
%
%
%
\def\unredoffs{} \def\redoffs{\voffset=-.31truein\hoffset=-.59truein}
\def\speclscape{\special{ps: landscape}}
%
%
%
%
\newbox\leftpage \newdimen\fullhsize \newdimen\hstitle \newdimen\hsbody
\tolerance=1000\hfuzz=2pt
\catcode`\@=11 
\def\bigans{b }
\def\answ{b }
\ifx\answ\bigans\message{(This will come out unreduced.}
\magnification=1200\unredoffs\baselineskip=16pt plus 2pt minus 1pt
\hsbody=\hsize \hstitle=\hsize 
\else\message{(This will be reduced.} \let\l@r=L
\magnification=1000\baselineskip=16pt plus 2pt minus 1pt \vsize=7truein
\redoffs \hstitle=8truein\hsbody=4.75truein\fullhsize=10truein\hsize=\hsbody
\output={\ifnum\pageno=0 
  \shipout\vbox{\speclscape{\hsize\fullhsize\makeheadline}
    \hbox to \fullhsize{\hfill\pagebody\hfill}}\advancepageno
  \else
  \almostshipout{\leftline{\vbox{\pagebody\makefootline}}}\advancepageno
  \fi}
\def\almostshipout#1{\if L\l@r \count1=1 \message{[\the\count0.\the\count1]}
      \global\setbox\leftpage=#1 \global\let\l@r=R
 \else \count1=2
  \shipout\vbox{\speclscape{\hsize\fullhsize\makeheadline}
      \hbox to\fullhsize{\box\leftpage\hfil#1}}  \global\let\l@r=L\fi}
\fi
%
\newcount\yearltd\yearltd=\year\advance\yearltd by -1900

\def\Title#1#2{\nopagenumbers\abstractfont\hsize=\hstitle\rightline{#1}%
\vskip 1in\centerline{\titlefont #2}\abstractfont\vskip .5in\pageno=0}
\def\Date#1{\vfill\leftline{#1}\tenpoint\supereject\global\hsize=\hsbody%
\footline={\hss\tenrm\folio\hss}}
%

\def\draftmode{\message{ DRAFTMODE }\def\draftdate{{\rm preliminary draft:
\number\month/\number\day/\number\yearltd\ \ \hourmin}}%
\headline={\hfil\draftdate}\writelabels\baselineskip=20pt plus 2pt minus 2pt
 {\count255=\time\divide\count255 by 60 \xdef\hourmin{\number\count255}
  \multiply\count255 by-60\advance\count255 by\time
  \xdef\hourmin{\hourmin:\ifnum\count255<10 0\fi\the\count255}}}
\def\nolabels{\def\wrlabeL##1{}\def\eqlabeL##1{}\def\reflabeL##1{}}
\def\writelabels{\def\wrlabeL##1{\leavevmode\vadjust{\rlap{\smash%
{\line{{\escapechar=` \hfill\rlap{\sevenrm\hskip.03in\string##1}}}}}}}%
\def\eqlabeL##1{{\escapechar-1\rlap{\sevenrm\hskip.05in\string##1}}}%
\def\reflabeL##1{\noexpand\llap{\noexpand\sevenrm\string\string\string##1}}}
\nolabels
%
\global\newcount\secno \global\secno=0
\global\newcount\meqno \global\meqno=1
\def\newsec#1{\global\advance\secno by1\message{(\the\secno. #1)}
\global\subsecno=0\eqnres@t\noindent{\bf\the\secno. #1}
\writetoca{{\secsym} {#1}}\par\nobreak\medskip\nobreak}
\def\eqnres@t{\xdef\secsym{\the\secno.}\global\meqno=1\bigbreak\bigskip}
\def\sequentialequations{\def\eqnres@t{\bigbreak}}\xdef\secsym{}
\global\newcount\subsecno \global\subsecno=0
\def\subsec#1{\global\advance\subsecno by1\message{(\secsym\the\subsecno.
#1)}
\ifnum\lastpenalty>9000\else\bigbreak\fi
\noindent{\it\secsym\the\subsecno. #1}\writetoca{\string\quad
{\secsym\the\subsecno.} {#1}}\par\nobreak\medskip\nobreak}
\def\appendix#1#2{\global\meqno=1\global\subsecno=0\xdef\secsym{\hbox{#1.}}
\bigbreak\bigskip\noindent{\bf Appendix #1. #2}\message{(#1. #2)}
\writetoca{Appendix {#1.} {#2}}\par\nobreak\medskip\nobreak}
%
%
\def\eqnn#1{\xdef #1{(\secsym\the\meqno)}\writedef{#1\leftbracket#1}%
\global\advance\meqno by1\wrlabeL#1}
\def\eqna#1{\xdef #1##1{\hbox{$(\secsym\the\meqno##1)$}}
\writedef{#1\numbersign1\leftbracket#1{\numbersign1}}%
\global\advance\meqno by1\wrlabeL{#1$\{\}$}}
\def\eqn#1#2{\xdef #1{(\secsym\the\meqno)}\writedef{#1\leftbracket#1}%
\global\advance\meqno by1$$#2\eqno#1\eqlabeL#1$$}
%
\newskip\footskip\footskip14pt plus 1pt minus 1pt 
\def\footnotefont{\ninepoint}\def\f@t#1{\footnotefont #1\@foot}
\def\f@@t{\baselineskip\footskip\bgroup\footnotefont\aftergroup\@foot\let\next}
\setbox\strutbox=\hbox{\vrule height9.5pt depth4.5pt width0pt}
\global\newcount\ftno \global\ftno=0
\def\foot{\global\advance\ftno by1\footnote{$^{\the\ftno}$}}
%
\newwrite\ftfile
\def\footend{\def\foot{\global\advance\ftno by1\chardef\wfile=\ftfile
$^{\the\ftno}$\ifnum\ftno=1\immediate\openout\ftfile=foots.tmp\fi%
\immediate\write\ftfile{\noexpand\smallskip%
\noexpand\item{f\the\ftno:\ }\pctsign}\findarg}%
\def\footatend{\vfill\eject\immediate\closeout\ftfile{\parindent=20pt
\centerline{\bf Footnotes}\nobreak\bigskip\input foots.tmp }}}
\def\footatend{}
%
%
\global\newcount\refno \global\refno=1
\newwrite\rfile
\def\ref{[\the\refno]\nref}
\def\nref#1{\xdef#1{[\the\refno]}\writedef{#1\leftbracket#1}%
\ifnum\refno=1\immediate\openout\rfile=refs.tmp\fi
\global\advance\refno by1\chardef\wfile=\rfile\immediate
\write\rfile{\noexpand\item{#1\ }\reflabeL{#1\hskip.31in}\pctsign}\findarg}
\def\findarg#1#{\begingroup\obeylines\newlinechar=`\^^M\pass@rg}
{\obeylines\gdef\pass@rg#1{\writ@line\relax #1^^M\hbox{}^^M}%
\gdef\writ@line#1^^M{\expandafter\toks0\expandafter{\striprel@x #1}%
\edef\next{\the\toks0}\ifx\next\em@rk\let\next=\endgroup\else\ifx\next\empty%
\else\immediate\write\wfile{\the\toks0}\fi\let\next=\writ@line\fi\next\relax}}
\def\striprel@x#1{} \def\em@rk{\hbox{}}
\def\lref{\begingroup\obeylines\lr@f}
\def\lr@f#1#2{\gdef#1{\ref#1{#2}}\endgroup\unskip}
\def\semi{;\hfil\break}
\def\addref#1{\immediate\write\rfile{\noexpand\item{}#1}} 
\def\startrefs#1{\immediate\openout\rfile=refs.tmp\refno=#1}
\def\xref{\expandafter\xr@f}\def\xr@f[#1]{#1}
\def\refs#1{\count255=1[\r@fs #1{\hbox{}}]}
\def\r@fs#1{\ifx\und@fined#1\message{reflabel \string#1 is undefined.}%
\nref#1{need to supply reference \string#1.}\fi%
\vphantom{\hphantom{#1}}\edef\next{#1}\ifx\next\em@rk\def\next{}%
\else\ifx\next#1\ifodd\count255\relax\xref#1\count255=0\fi%
\else#1\count255=1\fi\let\next=\r@fs\fi\next}
%

%
\newwrite\ffile\global\newcount\figno \global\figno=1
\def\fig{fig.~\the\figno\nfig}
\def\nfig#1{\xdef#1{fig.~\the\figno}%
\writedef{#1\leftbracket fig.\noexpand~\the\figno}%
\ifnum\figno=1\immediate\openout\ffile=figs.tmp\fi\chardef\wfile=\ffile%
\immediate\write\ffile{\noexpand\medskip\noexpand\item{Fig.\ \the\figno. }
\reflabeL{#1\hskip.55in}\pctsign}\global\advance\figno by1\findarg}
\def\xfig{\expandafter\xf@g}\def\xf@g fig.\penalty\@M\ {}
\def\figs#1{figs.~\f@gs #1{\hbox{}}}
\def\f@gs#1{\edef\next{#1}\ifx\next\em@rk\def\next{}\else
\ifx\next#1\xfig #1\else#1\fi\let\next=\f@gs\fi\next}
\newwrite\lfile
{\escapechar-1\xdef\pctsign{\string\%}\xdef\leftbracket{\string\{}
\xdef\rightbracket{\string\}}\xdef\numbersign{\string\#}}

\def\writestop{\def\writestoppt{\immediate\write\lfile{\string\pageno%
\the\pageno\string\startrefs\leftbracket\the\refno\rightbracket%
\string\def\string\secsym\leftbracket\secsym\rightbracket%
\string\secno\the\secno\string\meqno\the\meqno}\immediate\closeout\lfile}}
\def\writestoppt{}\def\writedef#1{}
\def\seclab#1{\xdef #1{\the\secno}\writedef{#1\leftbracket#1}\wrlabeL{#1=#1}}
\def\subseclab#1{\xdef #1{\secsym\the\subsecno}%
\writedef{#1\leftbracket#1}\wrlabeL{#1=#1}}
\newwrite\tfile \def\writetoca#1{}
\def\leaderfill{\leaders\hbox to 1em{\hss.\hss}\hfill}
\def\writetoc{\immediate\openout\tfile=toc.tmp
   \def\writetoca##1{{\edef\next{\write\tfile{\noindent ##1
   \string\leaderfill {\noexpand\number\pageno} \par}}\next}}}

%
\catcode`\@=12 
%
\edef\tfontsize{\ifx\answ\bigans scaled\magstep3\else scaled\magstep4\fi}
\font\titlerm=cmr10 \tfontsize \font\titlerms=cmr7 \tfontsize
\font\titlermss=cmr5 \tfontsize \font\titlei=cmmi10 \tfontsize
\font\titleis=cmmi7 \tfontsize \font\titleiss=cmmi5 \tfontsize
\font\titlesy=cmsy10 \tfontsize \font\titlesys=cmsy7 \tfontsize
\font\titlesyss=cmsy5 \tfontsize \font\titleit=cmti10 \tfontsize
\skewchar\titlei='177 \skewchar\titleis='177 \skewchar\titleiss='177
\skewchar\titlesy='60 \skewchar\titlesys='60 \skewchar\titlesyss='60
\def\titlefont{\def\rm{\fam0\titlerm}
\textfont0=\titlerm \scriptfont0=\titlerms \scriptscriptfont0=\titlermss
\textfont1=\titlei \scriptfont1=\titleis \scriptscriptfont1=\titleiss
\textfont2=\titlesy \scriptfont2=\titlesys \scriptscriptfont2=\titlesyss
\textfont\itfam=\titleit \def\it{\fam\itfam\titleit}\rm}
 \ifx\answ\bigans\else scaled\magstep1\fi
\ifx\answ\bigans\def\abstractfont{\tenpoint}\else
\font\abssl=cmsl10 scaled \magstep1
\font\absrm=cmr10 scaled\magstep1 \font\absrms=cmr7 scaled\magstep1
\font\absrmss=cmr5 scaled\magstep1 \font\absi=cmmi10 scaled\magstep1
\font\absis=cmmi7 scaled\magstep1 \font\absiss=cmmi5 scaled\magstep1
\font\abssy=cmsy10 scaled\magstep1 \font\abssys=cmsy7 scaled\magstep1
\font\abssyss=cmsy5 scaled\magstep1 \font\absbf=cmbx10 scaled\magstep1
\skewchar\absi='177 \skewchar\absis='177 \skewchar\absiss='177
\skewchar\abssy='60 \skewchar\abssys='60 \skewchar\abssyss='60
\def\abstractfont{\def\rm{\fam0\absrm}
\textfont0=\absrm \scriptfont0=\absrms \scriptscriptfont0=\absrmss
\textfont1=\absi \scriptfont1=\absis \scriptscriptfont1=\absiss
\textfont2=\abssy \scriptfont2=\abssys \scriptscriptfont2=\abssyss
\textfont\itfam=\bigit \def\it{\fam\itfam\bigit}\def\footnotefont{\tenpoint}%
\textfont\slfam=\abssl \def\sl{\fam\slfam\abssl}%
\textfont\bffam=\absbf \def\bf{\fam\bffam\absbf}\rm}\fi
\def\tenpoint{\def\rm{\fam0\tenrm}
\textfont0=\tenrm \scriptfont0=\sevenrm \scriptscriptfont0=\fiverm
\textfont1=\teni  \scriptfont1=\seveni  \scriptscriptfont1=\fivei
\textfont2=\tensy \scriptfont2=\sevensy \scriptscriptfont2=\fivesy
\textfont\itfam=\tenit
\def\it{\fam\itfam\tenit}\def\footnotefont{\ninepoint}%
\textfont\bffam=\tenbf \def\bf{\fam\bffam\tenbf}\def\sl{\fam\slfam\tensl}\rm}
\font\ninerm=cmr9 \font\sixrm=cmr6 \font\ninei=cmmi9 \font\sixi=cmmi6
\font\ninesy=cmsy9 \font\sixsy=cmsy6 \font\ninebf=cmbx9
\font\nineit=cmti9 \font\ninesl=cmsl9 \skewchar\ninei='177
\skewchar\sixi='177 \skewchar\ninesy='60 \skewchar\sixsy='60
\def\ninepoint{\def\rm{\fam0\ninerm}
\textfont0=\ninerm \scriptfont0=\sixrm \scriptscriptfont0=\fiverm
\textfont1=\ninei \scriptfont1=\sixi \scriptscriptfont1=\fivei
\textfont2=\ninesy \scriptfont2=\sixsy \scriptscriptfont2=\fivesy
\textfont\itfam=\ninei \def\it{\fam\itfam\nineit}\def\sl{\fam\slfam\ninesl}%
\textfont\bffam=\ninebf \def\bf{\fam\bffam\ninebf}\rm}
%
%

\hyphenation{anom-aly anom-alies coun-ter-term coun-ter-terms}
\def\inv{^{\raise.15ex\hbox{${\scriptscriptstyle -}$}\kern-.05em 1}}

\def\Dsl{\,\raise.15ex\hbox{/}\mkern-13.5mu D} 
\def\dsl{\raise.15ex\hbox{/}\kern-.57em\partial}

\font\bigit=cmti10 scaled \magstep1
\def\lspace{\ifx\answ\bigans{}\else\qquad\fi}
\def\lbspace{\ifx\answ\bigans{}\else\hskip-.2in\fi} 
\def\boxeqn#1{\vcenter{\vbox{\hrule\hbox{\vrule\kern3pt\vbox{\kern3pt
           \hbox{${\displaystyle #1}$}\kern3pt}\kern3pt\vrule}\hrule}}}
\def\mbox#1#2{\vcenter{\hrule \hbox{\vrule height#2in
               \kern#1in \vrule} \hrule}}  
%

\def\e#1{{\rm e}^{^{\textstyle#1}}}

\def\darr#1{\raise1.5ex\hbox{$\leftrightarrow$}\mkern-16.5mu #1}

\def\half{{\textstyle{1\over2}}} 
\def\roughly#1{\raise.3ex\hbox{$#1$\kern-.75em\lower1ex\hbox{$\sim$}}}



\def\IB{\relax\hbox{$\inbar\kern-.3em{\rm B}$}}
\def\IC{\relax\hbox{$\inbar\kern-.3em{\rm C}$}}
\def\ID{\relax\hbox{$\inbar\kern-.3em{\rm D}$}}
\def\IE{\relax\hbox{$\inbar\kern-.3em{\rm E}$}}
\def\IF{\relax\hbox{$\inbar\kern-.3em{\rm F}$}}
\def\IG{\relax\hbox{$\inbar\kern-.3em{\rm G}$}}
\def\IGa{\relax\hbox{${\rm I}\kern-.18em\Gamma$}}
\def\IH{\relax{\rm I\kern-.18em H}}
\def\IK{\relax{\rm I\kern-.18em K}}
\def\II{\relax{\rm I\kern-.18em I}}
\def\IL{\relax{\rm I\kern-.18em L}}
\def\IP{\relax{\rm I\kern-.18em P}}
\def\IR{\relax{\rm I\kern-.18em R}}
\def\IZ{\relax\ifmmode\mathchoice {\hbox{\cmss Z\kern-.4em Z}}{\hbox{\cmss
Z\kern-.4em Z}} {\lower.9pt\hbox{\cmsss Z\kern-.4em Z}}
{\lower1.2pt\hbox{\cmsss Z\kern-.4em Z}}\else{\cmss Z\kern-.4em Z}\fi}

\def\IB{\relax{\rm I\kern-.18em B}}
\def\IC{{\relax\hbox{$\inbar\kern-.3em{\rm C}$}}}
\def\ID{\relax{\rm I\kern-.18em D}}
\def\IE{\relax{\rm I\kern-.18em E}}
\def\IF{\relax{\rm I\kern-.18em F}}


\def\p{\partial}




\def\s{\lies}


\def\half {{1\over 2}}


\def\a{\alpha}
\def\b{\beta}
\def\g{\gamma}  \def\G{\Gamma}
\def\d{\delta}  
\def\m{\mu}
\def\n{\nu}

\def\l{\lambda} 
\def\k{\kappa}
\def\e{\epsilon}

\def\|{\Big|}
\def\({\Big(}   \def\){\Big)}
\def\[{\Big[}   \def\]{\Big]}







\def\unlockat{\catcode`\@=11}
\def\lockat{\catcode`\@=12}

\unlockat


\def\newsec#1{\global\advance\secno by1\message{(\the\secno. #1)}
\global\subsecno=0\global\subsubsecno=0\eqnres@t\noindent {\bf\the\secno. #1}
\writetoca{{\secsym} {#1}}\par\nobreak\medskip\nobreak}
\global\newcount\subsecno \global\subsecno=0
\def\subsec#1{\global\advance\subsecno by1\message{(\secsym\the\subsecno.
#1)}
\ifnum\lastpenalty>9000\else\bigbreak\fi\global\subsubsecno=0
\noindent{\it\secsym\the\subsecno. #1}
\writetoca{\string\quad {\secsym\the\subsecno.} {#1}}
\par\nobreak\medskip\nobreak}
\global\newcount\subsubsecno \global\subsubsecno=0
\def\subsubsec#1{\global\advance\subsubsecno by1
\message{(\secsym\the\subsecno.\the\subsubsecno. #1)}
\ifnum\lastpenalty>9000\else\bigbreak\fi
\noindent\quad{\secsym\the\subsecno.\the\subsubsecno.}{#1}
\writetoca{\string\qquad{\secsym\the\subsecno.\the\subsubsecno.}{#1}}
\par\nobreak\medskip\nobreak}

\def\subsubseclab#1{\DefWarn#1\xdef #1{\noexpand\hyperref{}{subsubsection}%
{\secsym\the\subsecno.\the\subsubsecno}%
{\secsym\the\subsecno.\the\subsubsecno}}%
\writedef{#1\leftbracket#1}\wrlabeL{#1=#1}}
\lockat

\def\dbend{\lower3.5pt\hbox{\manual\char127}}


\def\boxit#1{\vbox{\hrule\hbox{\vrule\kern8pt
\vbox{\hbox{\kern8pt}\hbox{\vbox{#1}}\hbox{\kern8pt}}
\kern8pt\vrule}\hrule}}

\def\mathboxit#1{\vbox{\hrule\hbox{\vrule\kern8pt\vbox{\kern8pt
\hbox{$\displaystyle #1$}\kern8pt}\kern8pt\vrule}\hrule}}

\overfullrule=0pt


\def\p{\partial}




\def\half {{1\over 2}}


\def\a{\alpha}
\def\b{\beta}
\def\g{\gamma}  \def\G{\Gamma}
\def\d{\delta}  
\def\m{\mu}
\def\n{\nu}

\def\l{\lambda} 
\def\k{\kappa}
\def\e{\epsilon}

\def\a{\alpha}
\def\b{\beta}
\def\d{\delta}

\def\m{\mu}
\def\n{\nu}

\def\s{\sigma}
\def\l{\lambda}

\def\k{\kappa}

\def\t{\theta}


\def\|{\Big|}
\def\({\Big(}   \def\){\Big)}
\def\[{\Big[}   \def\]{\Big]}







\def\newdate{11/7/2002} 
 
\def\a{\alpha} 
\def\b{\beta} 
\def\g{\gamma} 
\def\l{\lambda} 
\def\d{\delta} 
\def\e{\epsilon} 
\def\t{\theta} 
 
\def\s{\sigma}

\def\G{\Gamma}

\def\p{\partial} 
\def\half{{1\over 2}}

\def\pg{{\rm pg}}
\def\af{{\rm af}}
 
  
\Title{\vbox{\hbox{YITP-SB-02-59}}}   
{\vbox{   
\centerline{Yang-Mills Theory  as an Illustration} 
\vskip .2cm   
\centerline{of the Covariant Quantization of Superstrings}  
\vskip .2cm   
}}   
\medskip\centerline
{
P.~A.~Grassi$^{~a,}$\foot{pgrassi@insti.physics.sunysb.edu},  
G.~Policastro$^{~b,}$\foot{policast@cibslogin.sns.it},   
and   
P.~van~Nieuwenhuizen$^{~a,}$\foot{vannieu@insti.physics.sunysb.edu}
} 
\medskip   
\centerline{$^{(a)}$ {\it C.N. Yang Institute for Theoretical Physics,} }  
\centerline{\it State University of New York at Stony Brook,   
NY 11794-3840, USA}  
\vskip .3cm  
\centerline{$^{(b)}$ {\it 
DAMTP, Centre for Mathematical Sciences
} }  
\centerline{\it Wilberforce Road - Cambridge CB3 0WA, UK}  
  
\medskip  
\vskip  1.5cm  
\noindent  
We present a new approach to the quantization of the superstring. After a brief review of the 
classical Green-Schwarz formulation for the superstring and Berkovits' 
approach to its quantization based on pure spinors, we discuss our formulation without 
pure spinor constraints. In order to illustrate the ideas on which our work is based, 
we apply them to pure Yang-Mills theory. 
In the appendices, we include some background material for the Green-Schwarz and 
Berkovits formulations, in order that this presentation be self contained. 

\vskip 1cm

\centerline{\it 
Based on a talk given at the Third Sacharov Conference.}
 
\Date{\newdate}  
  
  
\lref\lrp{  
U.~Lindstr\"om, M.~Ro\v cek, and P.~van Nieuwenhuizen, in preparation.   
}  
  
\lref\pr{  
P. van Nieuwenhuizen, in {\it Supergravity `81},   
Proceedings First School on Supergravity, Cambridge University Press, 1982, page 165.  
}  
  
\lref\polc{  
J.~Polchinski,  
{\it String Theory. Vol. 1: An Introduction To The Bosonic String,}  
{\it String Theory. Vol. 2: Superstring Theory And Beyond,}  
{\it  Cambridge, UK: Univ. Pr. (1998) 531 p}.  
}  
  
\lref\GreenSchwarz{
M.~B.~Green and  J.~H.~Schwarz, {\it Covariant Description Of Superstrings,}   
Phys.\  Lett.\ {\bf B136} (1984) 367\semi 
M.~B.~Green and J.~H.~Schwarz,  
{\it Properties Of The Covariant Formulation Of Superstring Theories,}  
Nucl.\ Phys.\ {\bf B243} (1984) 285
}

\lref\superstring{  
M.~B.~Green and C.~M.~Hull, QMC/PH/89-7  
{\it Presented at Texas A and M Mtg. on String Theory, College  
  Station, TX, Mar 13-18, 1989}\semi  
R.~Kallosh and M.~Rakhmanov, Phys.\ Lett.\  {\bf B209} (1988) 233\semi  
U. ~Lindstr\"om, M.~Ro\v cek, W.~Siegel,   
P.~van Nieuwenhuizen and A.~E.~van de Ven, Phys. Lett. {\bf B224} (1989)   
285, Phys. Lett. {\bf B227}(1989) 87, and Phys. Lett. {\bf B228}(1989) 53;   
S.~J.~Gates, M.~T.~Grisaru, U.~Lindstr\"om, M.~Ro\v cek, W.~Siegel,   
P.~van Nieuwenhuizen and A.~E.~van de Ven,  
{\it Lorentz Covariant Quantization Of The Heterotic Superstring,}  
Phys.\ Lett.\  {\bf B225} (1989) 44;   
A.~Mikovic, M.~Rocek, W.~Siegel, P.~van Nieuwenhuizen, J.~Yamron and  
A.~E.~van de Ven, Phys.\ Lett.\  {\bf B235} (1990) 106;   
U.~Lindstr\"om, M.~Ro\v cek, W.~Siegel, P.~van Nieuwenhuizen and  
A.~E.~van de Ven,   
{\it Construction Of The Covariantly Quantized Heterotic Superstring,}  
Nucl.\ Phys.\  {\bf B330} (1990) 19 \semi  
F. Bastianelli, G. W. Delius and E. Laenen, Phys. \ Lett. \ {\bf  
  B229}, 223 (1989)\semi  
R.~Kallosh, Nucl.\ Phys.\ Proc.\ Suppl.\  {\bf 18B}  
  (1990) 180 \semi  
M.~B.~Green and C.~M.~Hull, Mod.\ Phys.\ Lett.\  {\bf A5} (1990) 1399\semi   
M.~B.~Green and C.~M.~Hull, Nucl.\ Phys.\  {\bf B344} (1990) 115\semi  
F.~Essler, E.~Laenen, W.~Siegel and J.~P.~Yamron, Phys.\ Lett.\  {\bf B254} (1991) 411\semi   
  F.~Essler, M.~Hatsuda, E.~Laenen, W.~Siegel, J.~P.~Yamron, T.~Kimura  
  and A.~Mikovic,   
  Nucl.\ Phys.\  {\bf B364} (1991) 67\semi   
J.~L.~Vazquez-Bello,  
  Int.\ J.\ Mod.\ Phys.\  {\bf A7} (1992) 4583\semi  
E. Bergshoeff, R. Kallosh and A. Van Proeyen, ``Superparticle  
  actions and gauge fixings'', Class.\ Quant.\ Grav {\bf 9}   
  (1992) 321\semi  
C.~M.~Hull and J.~Vazquez-Bello, Nucl.\ Phys.\  {\bf B416}, (1994) 173 [hep-th/9308022]\semi  
P.~A.~Grassi, G.~Policastro and M.~Porrati,  
{\it Covariant quantization of the Brink-Schwarz superparticle,}  
Nucl.\ Phys.\ B {\bf 606}, 380 (2001)  
[arXiv:hep-th/0009239].  
}  
  
\lref\bv{  
N. Berkovits and C. Vafa,  
{\it $N=4$ Topological Strings}, Nucl. Phys. B433 (1995) 123,   
hep-th/9407190.}  
  
\lref\fourreview{N. Berkovits,  {\it Covariant Quantization Of  
The Green-Schwarz Superstring In A Calabi-Yau Background,}  
Nucl. Phys. {\bf B431} (1994) 258, ``A New Description Of The Superstring,''  
Jorge Swieca Summer School 1995, p. 490, hep-th/9604123.}  
  
\lref\OoguriPS{  
H.~Ooguri, J.~Rahmfeld, H.~Robins and J.~Tannenhauser,  
{\it Holography in superspace,}  
JHEP {\bf 0007}, 045 (2000)  
[arXiv:hep-th/0007104].  
}  
  
\lref\bvw{  
N.~Berkovits, C.~Vafa and E.~Witten,  
{\it Conformal field theory of AdS background with Ramond-Ramond flux,}  
JHEP {\bf 9903}, 018 (1999)  
[arXiv:hep-th/9902098].  
}  
\lref\wittwi{  
E.~Witten,  
{\it An Interpretation Of Classical Yang-Mills Theory,}  
Phys.\ Lett.\ B {\bf 77}, 394 (1978);   
E.~Witten,  
{\it Twistor - Like Transform In Ten-Dimensions,}  
Nucl.\ Phys.\ B {\bf 266}, 245 (1986)}  
  
\lref\SYM{  
W.~Siegel,  
{\it Superfields In Higher Dimensional Space-Time,}  
Phys.\ Lett.\ B {\bf 80}, 220 (1979)\semi  
B.~E.~Nilsson,  
{\it Pure Spinors As Auxiliary Fields In The Ten-Dimensional   
Supersymmetric Yang-Mills Theory,}  
Class.\ Quant.\ Grav.\  {\bf 3}, L41 (1986);   
B.~E.~Nilsson,  
{\it Off-Shell Fields For The Ten-Dimensional Supersymmetric   
Yang-Mills Theory,} GOTEBORG-81-6\semi  
S.~J.~Gates and S.~Vashakidze,  
{\it On D = 10, N=1 Supersymmetry, Superspace Geometry And Superstring Effects,}  
Nucl.\ Phys.\ B {\bf 291}, 172 (1987)\semi  
M.~Cederwall, B.~E.~Nilsson and D.~Tsimpis,  
{\it The structure of maximally supersymmetric Yang-Mills theory:    
Constraining higher-order corrections,}  
JHEP {\bf 0106}, 034 (2001)  
[arXiv:hep-th/0102009];   
M.~Cederwall, B.~E.~Nilsson and D.~Tsimpis,  
{\it D = 10 superYang-Mills at O(alpha**2),}  
JHEP {\bf 0107}, 042 (2001)  
[arXiv:hep-th/0104236].  
}  
\lref\har{  
J.~P.~Harnad and S.~Shnider,  
{\it Constraints And Field Equations For Ten-Dimensional   
Super-Yang-Mills Theory,}  
Commun.\ Math.\ Phys.\  {\bf 106}, 183 (1986).  
}  
\lref\wie{  
P.~B.~Wiegmann,  
{\it Multivalued Functionals And Geometrical Approach   
For Quantization Of Relativistic Particles And Strings,}   
Nucl.\ Phys.\ B {\bf 323}, 311 (1989).  
}  
\lref\purespinors{\'E. Cartan, {\it Lecons sur la th\'eorie des spineurs},   
Hermann, Paris (1937)\semi  
C. Chevalley, {\it The algebraic theory of Spinors},   
Columbia Univ. Press., New York\semi  
 R. Penrose and W. Rindler,   
{\it Spinors and Space-Time}, Cambridge Univ. Press, Cambridge (1984)   
\semi  
P. Budinich and A. Trautman, {\it The spinorial chessboard}, Springer,   
New York (1989).  
}  
\lref\coset{  
P.~Furlan and R.~Raczka,  
{\it Nonlinear Spinor Representations,}  
J.\ Math.\ Phys.\  {\bf 26}, 3021 (1985)\semi  
A.~S.~Galperin, P.~S.~Howe and K.~S.~Stelle,  
{\it The Superparticle and the Lorentz group,}  
Nucl.\ Phys.\ B {\bf 368}, 248 (1992)  
[arXiv:hep-th/9201020].  
}  
  
%
%
\lref\GS{M.B. Green, J.H. Schwarz, and E. Witten, {\it Superstring Theory,}   
 vol. 1, chapter 5 (Cambridge U. Press, 1987).    
}  
\lref\carlip{S. Carlip,   
{\it Heterotic String Path Integrals with the Green-Schwarz   
Covariant Action}, Nucl. Phys. B284 (1987) 365 \semi R. Kallosh,   
{\it Quantization of Green-Schwarz Superstring}, Phys. Lett. B195 (1987) 369.}   
 \lref\john{G. Gilbert and   
D. Johnston, {\it Equivalence of the Kallosh and Carlip Quantizations   
of the Green-Schwarz Action for the Heterotic String}, Phys. Lett. B205   
(1988) 273.}   
\lref\csm{W. Siegel, {\it Classical Superstring Mechanics}, Nucl. Phys. B263 (1986)   
93\semi   
W.~Siegel, {\it Randomizing the Superstring}, Phys. Rev. D 50 (1994), 2799.  
}     
\lref\sok{E. Sokatchev, {\it   
Harmonic Superparticle}, Class. Quant. Grav. 4 (1987) 237\semi   
E.R. Nissimov and S.J. Pacheva, {\it Manifestly Super-Poincar\'e   
Covariant Quantization of the Green-Schwarz Superstring},   
Phys. Lett. B202 (1988) 325\semi   
R. Kallosh and M. Rakhmanov, {\it Covariant Quantization of the   
Green-Schwarz Superstring}, Phys. Lett. B209 (1988) 233.}    
\lref\many{S.J. Gates Jr, M.T. Grisaru,   
U. Lindstrom, M. Rocek, W. Siegel, P. van Nieuwenhuizen and   
A.E. van de Ven, {\it Lorentz-Covariant Quantization of the Heterotic   
Superstring}, Phys. Lett. B225 (1989) 44\semi   
R.E. Kallosh, {\it Covariant Quantization of Type IIA,B   
Green-Schwarz Superstring}, Phys. Lett. B225 (1989) 49\semi   
M.B. Green and C.M. Hull, {\it Covariant Quantum Mechanics of the   
Superstring}, Phys. Lett. B225 (1989) 57.}    
 \lref\fms{D. Friedan, E. Martinec and S. Shenker,   
{\it Conformal Invariance, Supersymmetry and String Theory},   
Nucl. Phys. B271 (1986) 93.}  
\lref\kawai{  
T.~Kawai,  
{\it Remarks On A Class Of BRST Operators,}  
Phys.\ Lett.\ B {\bf 168}, 355 (1986).}  
 \lref\ufive{N. Berkovits, {\it   
Quantization of the Superstring with Manifest U(5) Super-Poincar\'e   
Invariance}, Phys. Lett. B457 (1999) 94, hep-th/9902099.}    
\lref\BerkovitsRB{ N.~Berkovits,   
{\it Covariant quantization of the superparticle   
using pure spinors,} [hep-th/0105050].    
}

  
\lref\BerkovitsFE{  
N.~Berkovits,  
{\it Super-Poincar\'e covariant quantization of the superstring,}  
JHEP { 0004}, 018 (2000)  
[hep-th/0001035].  
}  
  
\lref\BerkovitsPH{  
N.~Berkovits and B.~C.~Vallilo,  
{\it Consistency of super-Poincar\'e covariant superstring tree amplitudes,}  
JHEP { 0007}, 015 (2000)  
[hep-th/0004171].  
}  
  
\lref\BerkovitsNN{  
N.~Berkovits,  
{\it Cohomology in the pure spinor formalism for the superstring,}  
JHEP { 0009}, 046 (2000)  
[hep-th/0006003].  
}  
  
\lref\BerkovitsWM{  
N.~Berkovits,  
{\it Covariant quantization of the superstring,}  
Int.\ J.\ Mod.\ Phys.\ A { 16}, 801 (2001)  
[hep-th/0008145].  
}  
  
\lref\BerkovitsYR{  
N.~Berkovits and O.~Chandia,  
{\it Superstring vertex operators in an AdS(5) x S(5) background,}  
Nucl.\ Phys.\ B {\bf 596}, 185 (2001)  
[hep-th/0009168].  
}  
\lref\BerkovitsZY{  
N.~Berkovits,  
{\it The Ten-dimensional Green-Schwarz   
superstring is a twisted Neveu-Schwarz-Ramond string,}  
Nucl.\ Phys.\ B {\bf 420}, 332 (1994)  
[arXiv:hep-th/9308129].  
}  
  
\lref\BerkovitsUS{  
N.~Berkovits,  
{\it Relating the RNS and pure spinor formalisms for the superstring,}  
hep-th/0104247.  
}  
  
\lref\BerkovitsMX{  
N.~Berkovits and O.~Chandia,  
{\it Lorentz invariance of the pure spinor BRST cohomology   
for the  superstring,}  
hep-th/0105149.  
}  
  
\lref\GrassiUG{  
P.~A.~Grassi, G.~Policastro, M.~Porrati and P.~van Nieuwenhuizen,  
{\it Toward Covariant quantization of superstrings without pure spinor constraints},   
hep-th/0112162, to appear in JHEP.  
}  
  
\lref\GrassiSP{  
P.~A.~Grassi, G.~Policastro, M.~Porrati and P.~van Nieuwenhuizen,  
{\it The massles spectrum of covariant superstrings},   
hep-th/0202123, to appear in JHEP. 
}

\lref\GrassiXV{
P.~A.~Grassi, G.~Policastro and P.~van Nieuwenhuizen,
{\it On the BRST cohomology of superstrings with / without pure spinors,}
arXiv:hep-th/0206216.
}
  
\lref\GrassiZZ{  
P.~A.~Grassi, A. Iglesias, G.~Policastro, and P.~van Nieuwenhuizen,  
{\it The Covariant Quantum Superstring and Superparticle from their calssical  
action},   
hep-th/020.  
}

\lref\WittenZZ{  
E.~Witten,  
{\it Mirror manifolds and topological field theory,}  
hep-th/9112056.  
}  
  
\lref\wichen{  
E.~Witten,  
{\it Chern-Simons gauge theory as a string theory,}  
arXiv:hep-th/9207094.  
}  
  
  
\lref\howe{P.S. Howe, {\it Pure Spinor Lines in Superspace and   
Ten-Dimensional Supersymmetric Theories},   
Phys. Lett. B258 (1991) 141, Addendum-ibid.B259 (1991) 511\semi   
P.S. Howe, {\it Pure Spinors, Function Superspaces and Supergravity   
Theories in Ten Dimensions and Eleven Dimensions}, Phys. Lett. B273 (1991)   
90.}  
  
\lref\tonin{I. Oda and M. Tonin, {\it On the Berkovits covariant quantization   
of the GS superstring},   
Phys. Lett. B520 (2001) 398 [hep-th/0109051]\semi   
M.~Matone, L.~Mazzucato, I.~Oda, D.~Sorokin and M.~Tonin,
{ \it 
The superembedding origin of the Berkovits pure spinor covariant  quantization of superstrings,}
arXiv:hep-th/0206104.
}  
  
\lref\Membrane{  
N.~Berkovits,  
{\it Towards covariant quantization of the supermembrane,}  
arXiv:hep-th/0201151.  
}  
  
\lref\MMOST{
}

\lref\grr{
M.B. Green,
Phys. Lett. B {\bf 223} (1989) 157
} 

\lref\GrassiXF{
P.~A.~Grassi, G.~Policastro and P.~van Nieuwenhuizen,
{\it The covariant quantum superstring and superparticle from their  classical actions,}
arXiv:hep-th/0209026.
}

\lref\berko{
N.~Berkovits,
{\it Super-Poincar\'e covariant quantization of the superstring,}
JHEP { 0004}, 018 (2000)
[hep-th/0001035]
\semi
N.~Berkovits and B.~C.~Vallilo,
{\it Consistency of super-Poincar\'e covariant superstring tree amplitudes,}
JHEP { 0007}, 015 (2000)
[hep-th/0004171]
\semi
N.~Berkovits,
{\it Cohomology in the pure spinor formalism for the superstring,}
JHEP { 0009}, 046 (2000)
[hep-th/0006003]
\semi
N.~Berkovits,
{\it Covariant quantization of the superstring,}
Int.\ J.\ Mod.\ Phys.\ A { 16}, 801 (2001)
[hep-th/0008145]
\semi
N.~Berkovits and O.~Chandia,
{\it Superstring vertex operators in an AdS(5) x S(5) background,}
Nucl.\ Phys.\ B {\bf 596}, 185 (2001)
[hep-th/0009168]
\semi
N.~Berkovits,
{\it The Ten-dimensional Green-Schwarz 
superstring is a twisted Neveu-Schwarz-Ramond string,}
Nucl.\ Phys.\ B {\bf 420}, 332 (1994)
[hep-th/9308129]
\semi
N.~Berkovits,
{\it Relating the RNS and pure spinor formalisms for the superstring,}
[hep-th/0104247]
\semi
N.~Berkovits and O.~Chandia,
{\it Lorentz invariance of the pure spinor BRST cohomology 
for the  superstring,}
[hep-th/0105149].
}


\baselineskip16pt  

\newsec{Introduction}

String theory is mostly based on the Ramond-Neveu-Schwarz (RNS)
formulation, with worldsheet fermions $\psi^m$ in the vector
representation of the spacetime Lorentz group $SO(9,1)$. This
formulation exhibits classically a $N=1$ local supersymmetry of the worldsheet.
The BRST symmetry of the RNS formulation is based on 
the super-reparametrization invariance of the worldsheet. The fundamental fields 
are the bosons $x^m$, the fermions $\psi^m$, 
the reparametrization ghosts $b_{zz}, c^z$ and the superghosts
$\beta, \g$. Physical states correspond to vertex operators which {i)} 
belong to the BRST cohomology and {ii)} are annihilated by $b_0$ for the 
open string, or by $b_0$ and $\tilde b_0$ for the closed string. 
To obtain a set of physical states which form a representation of spacetime supersymmetry, the GSO projection 
is applied to remove half of the physical states. Spacetime supersymmetry
is thus not manifest, and the study of Ramond-Ramond backgrounds is not
feasible. Therefore, one would prefer a formulation with spacetime
fermions $\t^\a$ belonging to a representation of $Spin(9,1)$ because
it would keep spacetime supersymmetry (susy) manifest. At the classical level, such a formulation was
constructed by Green and Schwarz in 1984 \GreenSchwarz. Their classical action
contains two fermions $\t^{i\a}$'s ($i=1,2$) and the bosonic coordinates
$x^m$. Each of the $\t$'s is real and can be chiral or anti-chiral (type IIA/B superstrings):
they are $16$-component Majorana-Weyl spinors which are spacetime
spinors and worldsheet scalars. We shall denote chiral spinors by
contravariant indices $\t^\a$ with $\a=1,\dots,16$; antichiral spinors
are denoted by $\t_\a$, also with $\a=1,\dots,16$.

The rigid spacetime supersymmetry is given by the usual non-linear
coordinate representation
\eqn\susyA{\d_\e \t^{i\a} = \e^{i\a}\,, ~~~~~ 
\d_\e x^m = \bar\e^i \G^m \t^i = i\, \e^{i \a} \g^m_{\a\b} \t^{i \b}\,,}
where $\g^m_{\a\b}$ are  real symmetric $16\times16$ matrices and the 
flavor indices $i=1,2$ are summed over. (In appendix A, Dirac matrices and 
Majorana-Weyl spinors are reviewed). 
Susy-invariant building blocks are 
\eqn\susyB{\Pi^m_\mu \equiv \p_\mu x^m - 
i \sum^2_{i=1} \t^i \g^m \p_\mu \t^i\,, ~~~~~ \p_\mu \t^{i\a}\,} where $\mu=0,1$
and $\p_0 = \p_t$ and $\p_1 = \p_\s$. A natural choice for the action
on a flat background spacetime and curved worldsheet would seem to be
\eqn\susyC{{\cal L} = {1\over 2 \pi} \, \sqrt{-h} \, 
h^{\mu\nu} \eta_{mn} \, \Pi^m_\mu \Pi^n_\nu\,, } 
with $h^{\m\n}$ the worldsheet metric, 
because it is the susy-invariant line element (a generalization of the
action for the bosonic string). However, it yields no kinetic term for the fermions. Even if 
one could produce a kinetic term, there would still be the problem that one would have
${1\over 2}(16+16) = 16$ fermionic propagating modes and $8$ bosonic
propagating modes. Such a theory could not yield a linear representation of supersymmetry.

A resolution of this problem became possible when Siegel found a new
local fermionic symmetry ($\k$-symmetry) for the point particle 
\lref\SiegelHH{
W.~Siegel,
{\it Hidden Local Supersymmetry In The Supersymmetric Particle Action,}
Phys.\ Lett.\ B {\bf 128}, 397 (1983).
}
\SiegelHH. 
Green and Schwarz tried to find this symmetry in their string, and they discovered
that it is present, but only after adding a Wess-Zumino-Novikov-Witten
term to the action. Using this symmetry one could impose the gauge
$\G^+ \t^1 = \G^+ \t^2 =0$ (where $\G^\pm = \G^0 \pm \G^9$), and if
one then also fixed the local scale and general coordinate symmetry by
$h^{\mu \nu } = \eta^{\mu\nu}$, and the remaining conformal symmetry by
$x^+(\s,t) = x^+_0 + p^+ t$, the action became a free string theory with $8$
fermionic degrees of freedom and $8$ bosonic degrees of freedom. Susy
was linearly realized and quantization posed no problem.

However, in this combined $\k$-light cone gauge, manifest $SO(9,1)$
Lorentz invariance is lost, and with it all the reasons for studying the
superstring in the first place. (We shall call the string of
Green and Schwarz the superstring, to distinguish it from the RNS
string which we call the spinning string.)

Going back to the original classical action, it was soon
realized that second class constraints were present, due to the
definition of the conjugate momenta of the $\t$'s. These second class
constraints could be handled by decomposing them w.r.t. a non-compact
$SU(5)$ subgroup of $SO(9,1)$ (see appendix D) , but then again
manifest Lorentz invariance was lost. An approach to quantization
which could deal with second class constraints and keep covariance 
was needed. By using a
proposal of Faddeev and Fradkin to add further fields, one could
turn second class constraints into first class constraints, but upon
quantization one now obtained an infinite set of ghosts-for-ghosts, and 
problems with the calculation of anomalies were encountered. At the end
of the 80's, several authors tried different approaches, but they always
encountered infinite sets of ghosts-for-ghosts, and 15 years of pain
followed \superstring.

A few years ago Berkovits developed a new line of thought \berko. 
Taking a flat background and a flat worldsheet metric, 
the central charge $c$ of 10 free 
bosons $x^m$ and one $\t$ is 
$c =10 - 2 \times 16 =-22$ 
(there is a conjugate momentum $p_{z\a}$ for $\t^\a$). He noted
that if one decomposes a chiral spinor $\l^\a$ under the
non-compact $SU(5)$ subgroup of $SO(9,1)$, it decomposes as
$\underline{16} \rightarrow
\underline{10}+\underline{5}^*+\underline{1}$ (see Appendix D). 
Imposing the constraint
\eqn\pure{\l^T \g^m \l = 0\,,}
also known as {\it pure spinor constraint}, 
one can express the $\underline{5}^*$ in terms of the $\underline{10}$
and $\underline{1}$, and hence it seemed that by adding a commuting
pure spinor (with conjugate momenta for the $\underline{10}$
and $\underline{1}$), one could obtain vanishing central charge: 
$c = 10_x - 2 \times 16_{\t,p_\t} + 2 \times (10 +1)_{\l, p_\l} = 0$. In
the past few years, he has developed this approach further.

Having a constraint such as \pure~in a theory
leads to problems at the quantum level in the computation of loop
corrections and in the definition of path integral. A similar situation 
occurred in superspace formulation of supergravity, where one must impose constraints 
on the supertorsions; in that case the constraints were solved and the 
covariance was sacrificed. One could work only with $\underline{10}$ and
$\underline{1}$, but then one would again violate manifest Lorentz
invariance.

We have developed an approach \GrassiUG~which starts with the same 
$\t^\a, p_{z\a}$
and $\l^\a$ as used by Berkovits, but we relax the constraint \pure~by
adding new ghosts. In Berkovits' and our approach one has the BRST
law $s\, \t^\a = i \l^\a$, with real $\t^\a$, but in Berkovits' approach $\l^\a$ 
must be complex in order that \pure~have a solution at all, 
{\it whereas in our approach $\l^\a$ is real}. 
The law $s \t^\a = i\, \l^\a$ is an enormous simplification over the law one would 
obtain from the $\kappa$-symmetry law $\delta_\kappa \t^\a = \Pi^m_\mu (\g_m \k^\mu)^\a$ 
with selfdual $\k^{\mu\a}$. It is this simpler starting point that avoids the infinite set of 
ghosts-for-ghosts. 
First, we give a brief review of the classical superstring action from which 
we shall only extract a set of first class constraints $d_{z\a}$. These first 
class constraints are removed from the action and used to construct a BRST charge. 

We deduce the full theory 
by requiring nilpotency of the BRST charge: each time nilpotency on a given field 
does not hold we add a new field (ghost) and define its BRST transformation 
rule such that nilpotency holds.  A priori, one might expect that one would end up again with an 
infinite set of ghosts-for-ghosts, but to our happy surprise the iteration procedure 
stops after a finite number of steps. 

In some modern approaches the difference between the action and the BRST charge becomes 
less clear (in the BV formalism the action is even equal to the BRST charge). So the 
transplantation of the first class constraints from the action to the BRST charge may not be as 
drastic as it may sound at first. We may in this way create a different off-shell formulation of the 
same physical theory. The great advantage of this procedure is that one is left with a free 
action, so that propagators become very easy to write down, and OPE's among vertex operators 
become as easy as in the RSN approach.  

We shall now present our approach. We have a new definition of physical states, 
and we obtain the correct spectrum for the open string as well as for the closed 
superstring, both at the massless level and at the massive levels. Since these 
notes are intended as introduction to our work, we give much background material 
in the appendices. Such material is not present in our papers, but 
may help to understand the reasons and the technical aspects of our approach. 

We have found since the conference some deep geometrical meanings of the new ghosts, but we have 
not yet found the underlying classical action to which our quantum theory corresponds. Sorokin, Tonin and 
collaborators have recently shown \tonin~how one can obtain Berkovits' theory 
from a $N=(2,0)$ worldsheet action with superdiffeomorphism embeddings, and it is possible 
that a similar approach yields our theory. 

\newsec{The classical Green-Schwarz action}

As we already mentioned, a natural generalization of the bosonic string with ${\cal L} \sim (\p_\a x^m)^2$ with 
spacetime supersymmetry is the supersymmetric line element given in \susyB~and \susyC. 
If one considers the interaction term $\p_\mu x^m (\t \g_m \p^\m \t)$ and if 
one chooses the light cone gauge $x^+ = x^+_0 + p^+ t$ one obtains a term 
$p^+ \t \g_+ \p_t \t= (\sqrt{p^+} \t) \g_+ \p_t (\sqrt{p^+} \t)$. This is not a satisfactory 
kinetic term because we also would need a term with $ p^+ \t \g_+ \p_\s \t$. Such a term 
would be obtained if the action contains a term of the form $ (\p_t x^+) \t \g_+ \p_\s \t$, or in covariant 
notation $\e^{\m\n} (\p_\m x^m) \t \g_m \p_\n \t$. 
The extra kinetic term  $\e^{\m\n} \p_\m x^m \t \g_m \p_\n \t$ is part of a Wess-Zumino term, 
(see appendix B).

Rigid susy \susyA~and $\d_\e (\p_\s x^+) = 0$ 
would lead to $\e \g^+ \p_\s \t =0$. This suggests that the light-cone gauge for $\t$ 
should read $\g^+ \t =0$. Since $\g^+ \t =0$, also $\t \g^+=0$, and using $\{\g^+, \g^-\} = 1$, one would also 
find that $\t \g^I \p_\s \t =0$ for $I=1, \dots, 8$. So, then we would find in the light cone gauge 
that the action for $\t$ becomes a free action, a good starting point for string theory at the 
quantum level. 

In order that these steps are correct, we would need a local fermionic symmetry which would justify 
the gauge $\g^+ \t =0$.  Pursuing this line of thought, one arrives then at the 
crucial question: does the sum of the supersymmetric line element and the WZNW term contain a new fermionic 
symmetry with half as many parameters as there are $\t$ components ? The answer is affirmative, and the 
$\k$-symmetry is briefly discussed at the end of appendix B, but since we shall not need the explicit form of 
the $\kappa$ symmetry transformation laws, we do not give them.  

The superstring action is very complicated already in a flat background. We extract from it a set of 
first class constraints $d_{z\a}= 0$, from which we build the BRST charge, and at all stages we work with a 
free action. 
The precise way to obtain $d_{z\a}$ from the classical superstring action is discussed in appendix B. 


\newsec{Determining the theory from  the nilpotency the BRST charge}

We now start our program of determining the theory 
the BRST charge and the ghost content) by requiring 
nilpotency of the BRST transformations. We consider only $\t$ for 
simplicity (we have also extended 
our work to two $\t$'s. We shall be careful (for once) with aspects such as reality and normalizations. 
The BRST transformations preserve reality and are generated by $\Lambda Q$ where $\Lambda$ 
is imaginary and anti-commuting. It then follows that $Q$ should also be antihermitian in order that 
$\Lambda\, Q$ be antihermitian. For any field, we define 
the $s$ transformations as BRST transformations without $\Lambda$, 
so $\d_B \Phi = [\Lambda Q, \Phi]$ and $s \Phi = [Q, \Phi]_{\pm}$. The $s$-transformations  have reality properties 
which follows from the BRST trasnformations (which preserve reality). 

We begin with 
\eqn\BRSTA{Q = \int i \l^\a d_{z\a}\,,}
where $d_{z\a}$ is given in Appendix C and 
$\int = {1\over 2 \pi} \oint dz$, which is indeed antihermitian because $d_{z \a}$ is 
antihermitian. (We have performed a Wick rotation in appendix C, in order to be able to use the conventional tools 
of conformal field theory, but the reality properties hold in Minkowski space).
The BRST operator depends on Heisenberg fields which satisfy the field 
equations, and since we work with a free action, $\bar\p \l^\a =0$ and $\bar\p d_{z\a} =0$ so that
in flat space $\l^\a d_{z \a}$ is a holomorphic current, namely 
$\bar\p (\l^\a d_{z \a}) = 0$. 

The field $d_{z\a}$ contains a term $p_{z\a}$, where $p_{z\a}$ is the 
momentum conjugate to $\t^\a$ and it is antihermitian since $p_{z\a}$ 
is antihermitian as can be seen from the action $\int d^2z p_{z\a} \bar\p \t$. The factor ${1\over2}$ in  $d_{z\a}$ 
is fixed by requiring that the OPE\foot{The OPE of $d_\a$ with $d_\b$ is evaluated 
using $\p x^m(z) \p x^n(w) \sim - \eta^{mn} (z-w)^{-2}$ and $p_{z\a}(z) \t^\b (w) \sim \d_\a^{~~\b} (z-w)^{-1}.$} 
of $d_\a$ with $d_\b$ be proportional to $\Pi^m_w$. The expression for $\Pi^m_z$ is real and fixed by spacetime 
susy. 

The operators $d_{z\a}$ generate a closed algebra of current with a central charge
\eqn\BRSTB{
d_\a(z) d_\b(w) \sim  2 i {{\g^m_{\a\b}\Pi_m(w)}\over{z-w}}\,,
~~~~~~~
d_\a(z) \Pi^m(w) \sim - 2 i {{\g^m_{\a\b}\p\t^\b(w)} \over {z-w}}\,, 
}
$$
\Pi^m(z) \Pi^n(w) \sim - {1\over (z-w)^2} \eta^{mn}\,,~~~~~~~
	d_\a(z) \t^\b(w) \sim {1\over {z-w}} \delta^{~\b}_{\a}\,. 
$$

Acting with \BRSTA~on $\t^\a$, one obtains 
$s \t^\a = i \l^\a$, and acting on $\l^\a$  yields $s \l^\a =0$. Nilpotency 
on $\t^\a$ and $\l^\a$ is achieved. 
Repeating this procedure on $x^m$
gives $s x^m = \l \g^m \t$, but since $s^2 x^m = i \l \g^m \l$
does not vanish, we introduce a new ghost $\xi^m$ by setting 
$s\,  x^m = \l \g^m \t + \xi^m$ and choosing the BRST transformation law of 
$\xi^m$ such that the nilpotency on $x^m$ is obtained. This leads to 
$s \xi^m = - i \l \g^m\l$. Nilpotency on $x^m$ is now achieved, but $s$ 
has acquired an extra
term\foot{Spacetime susy requires that $Q'$ depends on $\Pi^m_z$ 
instead of, for example $\p_z x^m$.} 
$Q' = - \int \xi^m \Pi_{z m}$ where we recall $\Pi^m_z  = 
\p_z x^m - i \t \g^m \p_z \t$. Nilpotency on $p_{z\a}$, or
equivalently on $d_{z\a}$, is obtained by further 
modifying the sum of 
$Q d_{z\a} = 
- 2 \Pi^m_z (\g_m \l)_\a$ and 
$ Q' d_{z \a} = - 2 i \xi^m (\g_m \p_z \t)_\a$ 
by adding 
$Q''  d_{z \a} = 
\p_z\chi_\a$ and fixing the BRST law of $\chi_\a$ such that nilpotency on
$d_{z\a}$ is achieved.\foot{Since $(Q+Q') d_{z\a} = \p_z( -2 \xi^m \g_m \l) _\a$, 
we add a term $\p_z \chi_\a$ instead of a field $\chi_{z\a}$.} 
This yields $Q \chi_\a = 2\xi^m (\g_m \l)_\a$
and $Q^2 \chi_\a =0$ due to a Fierz rearrangement involving three 
chiral spinors. At this point we have achieved nilpotency on $\t^\a,
x^m, d_{z\a}$ and $\l^\a, \xi^m, \chi_\a$. We introduce the antighosts
$w_{z\a}, \b_{z m}, \k^\a_z$ for the ghosts $\l^\a, \xi^m, \chi_\a$ and
find that $s \Phi = [Q, \Phi\}$ with
\eqn\BRST{
Q = \oint \Big(i\l^\a d_{z\a} -\xi^m \Pi_{z m} - \chi_\a \p_z \t^\a 
- 2 \xi^m (\k_z \g_m \l) -i \b_{z m} \l\g^m \l\Big)
}
reproduces all BRST laws obtained so far. 

Unfortunately, the BRST charge \BRST~fails to be nilpotent and 
therefore the concept of the BRST cohomology is at this point 
meaningless. In order to repair this problem, we could proceed in two different 
ways: {\it i)} either continuing with our program and requiring nilpotency on each field 
separately (on the antighosts $\beta_{zm}$, $\kappa_z^\a$ and $w_\a$); or {\it ii)} 
terminate this process by hand in one stroke by adding a ghost 
pair $(b,c_z)$ as we now explain. 
We begin with  
\eqn\nilA{
Q^2 = \int A_z \,,~~~~~~~ A_z = \xi_m \partial_z \xi^m + i  \lambda^\a \partial_z \chi_\a
- i \chi_\a \partial_z \lambda^\a \,.}
The non-closure term $A_z$ is due to the double poles 
in \BRSTB. By direct computation we establish that the anomaly $\int A_z$ is 
BRST invariant, as it should be according to consistency, 
$[Q,A_z] = \partial_z Y$ where $Y = {i \over 2} \xi_m \lambda \g^m \lambda$. 
If we define
\eqn\nilB{
Q' = Q + \int \left( c_z - {1\over 2} b B_z\right)\,,
}
with an hermitian $c_z$ and an antihermitian $b$, 
we find that 
\eqn\nilC{
{Q'}^2 = \int \Big( (A_z - B_z) +  {1\over 2} b [Q,B_z] \Big) \,,
}
and, requiring that $Q'$ be nilpotent, a solution for $B_z$ is obtained by imposing
\eqn\nilD{
[Q,B_z]=0 \,, ~~~~~ B_z = A_z + \partial X \,,~~~~~ [Q,X] = - Y
}
which is satisfied by $X=- {i\over 2} \chi_\a \lambda^\a$. Then one gets 
\eqn\nilE{
B_z = \xi_m \partial_z \xi^m + {1\over 2} \lambda^\a \partial_z \chi_\a - {3\over 2} 
\chi_\a \partial_z \lambda^\a\,.
}

However, any $Q'$ of the form $\int c_z +$``more'' can be always  brought in the form 
$\int c_z$ by a similarity transformations choosing the term denoted by ``more'' appropriately, 
namely as follows 
\eqn\newup{
Q' = \Big[ e^{{1\over 2}\int (- R_z - b\, S_z - b \p_z b\, T)} \int c_z 
e^{{1\over 2}\int (R_z + b\, S_z + b \p_z b\, T)}
\Big] 
}
$$ 
=\int \left(c_z + S_z - b \p_z T \right) +  
\Big[\int \left( S_z - b \p_z T \right), {\cal U} \Big] + 
{1\over 2}  \Big[ \Big[ \int \left(  S_z - b \p_z T \right), {\cal U} \Big],  {\cal U} \Big] + \dots
$$
where $ {\cal U} = \int (R_z + b\, S_z + b \p_z b\, T)$. The $R_z, S_z$ and 
$T$ are hermitian polynomials in all fields except $c_z,b$ with ghost numbers 
$0,1,2$, respectively.  The solution in \nilB~and \nilE~corresponds to a particular 
choice of $R_z, S_z$ and $T$, but any other choice also yields a nilpotent BRST 
charge. The operator $Q' = e^{-{\cal U}} \int c_z  e^{{\cal U}}$ has trivial  
cohomology in the space of local vertex operators with vanishing conformal 
spin, because any ${\cal O}(w)$ satisfying ${\int c_z} {\cal O}(w) =0$ can always be written 
as ${\cal O}(w) = b_0 {\cal G}(w)$ where ${\cal G}(w) = \int c_z {\cal O}(w)$. 
(Note that ${\cal O}(w)$ cannot depend on $c_z$, and $c_0 = \int c_z$). 

We shall restrict 
the space of vertex operators in which $Q$ acts, in order to obtain non-trivial 
cohomology. We achieve this by introducing a new quantum number, called grading, 
and requiring that vertex operators have non-negative grading. In the 
smaller space of non-negative grading the similarity transformation 
cannot transform each $Q$ into the form $\int c_z $, and we shall indeed 
obtain non-trivial cohomology, namely the correct cohomology. 

We have at this point obtained a nilpotent BRST charge, and a set of ghost (and 
antighost) fields (whose geometrical meaning at this point is becoming clear). It is time 
to revert to the issue of the central charge. Since all fields are free 
fields, one simply needs to add the central charge of each canonical pair: 
$c = 20$. So the central charge does not vanish, and to remedy this obstruction, we 
add by hand an anticommuting vector pair $(\omega^m,\eta^m_z)$ which contributes 
$- 2 \times 10 $ to $c$. The BRST charge does not contain $\omega^m$ and $\eta^m_z$, 
hence  $\omega^m$ and $\eta^m_z$ are BRST inert. 

The reader (and the authors) may feel uncomfortable with these 
rescue missions by hand, a good theory should produce all fields automatically 
without outside help. Fortunately, we can announce that a more 
fundamental way of proceeding, by  continuing to require 
nilpotency on the antighosts and then on the new fields which are introduced 
 in this process, produces the pair $(\omega^m,\eta^m_z)$! We are in the process of writing 
these consideration up, and hopefully also the pair $(b,c_z)$ will be automatically 
produced in this way. 

Our results obtained by elementary methods and ad hoc addition, display nevertheless 
a few striking regularities, which confirm us in our belief that we are on the right track. 

\newsec{The notion of the grading}

In our work we define physical states by means of vertex operators which 
satisfy two conditions

{\it i)} They are in the BRST cohomology

{\it ii)} They should have non-negative grading \GrassiSP. 

The grading is a quantum number which was initially obtained from the algebra of the abstract currents 
$d_{z\a}, \Pi^m_z$ and $\p_z \t^\a$. Assigning grading $-1$ to $d_{z\a}$, we assign grading $+1$ 
to the corresponding ghost $\l^\a$. We then require that the grading be preserved in the 
operator product expansion. From $d d \sim \Pi$ we deduce that $\Pi^m_z$ has grading $-2$, 
so $\xi^m$ has grading $2$. Then $d \Pi \sim \p\t$ assigns grading $-3$ to $\p\t$, and thus grading $+3$ 
to $\chi$. The grading of the ghosts $b$ and $c$ is more subtle, but it can be obtained in the same 
spirit. From $d \p\t \sim (z-w)^{-2}$ and $\Pi \Pi \sim (z-w)^{-2}$ we introduce 
a central charge generator $I$ which has grading $-4$. The corresponding 
ghost $c_z$ has grading $4$. All antighosts have opposite grading from the ghosts. The trivial 
ghost pair $\omega^m, \eta^m_z$ has grading $(4,-4)$ because it is part of a quartet of which the grading 
of the other members is already known \GrassiSP. 
With these grading assignements to the ghost fields, the BRST charge can be 
decomposed into pieces of non-negative grading $Q = \sum_{n=0}^{4} Q_{n}$ and 
it maps the subspace of the Hilbert space with non-negative grading into itself. In 
\GrassiXV, the equivalence with Berkovits' pure spinor formulation has been 
proven. 

According to the grading condition {\it ii)}, the most general expression for the massless 
vertex in the case of open superstring is given by 
\eqn\graddA{
{\cal O} = \l^\a A_\a + \xi^m A_m + \chi_\a W^\a + b-{\rm terms}\,
}
where $A_\a, A_m$ and $W^\a$ are arbitrary superfields, so $A_\a = A_\a(x,\t)$, etc..
Requiring non-negative grading, the following combinations 
\eqn\gradB{
b \l^\a \l^\b\,, ~~~~~~~ b \l^\a \xi^m\,,
}
are not allowed. Finally, requiring the BRST invariance, one easily derives 
the equations of motion for $N=1$ SYM in $D=(9,1)$. Along the 
same lines, one can study the closed string or massive operators and one finds the 
complete correct spectrum of the open or closed superstring. 

The notion that one must restrict the space of the vertex operators is not new 
by itself: in the spinning (RNS) string, one should restrict the commuting susy ghosts 
to non-negative mode numbers 
\lref\big{
M.~Henneaux,
{\it Brst Cohomology Of The Fermionic String,}
Phys.\ Lett.\ B {\bf 183}, 59 (1987); W.~Siegel,
{\it Boundary Conditions In First Quantization,}
Int.\ J.\ Mod.\ Phys.\ A {\bf 6}, 3997 (1991); 
N.~Berkovits, M.~T.~Hatsuda and W.~Siegel,
{\it The Big picture}
Nucl.\ Phys.\ B {\bf 371}, 434 (1992)
[arXiv:hep-th/9108021]} 
\big, and also in the bosonic string one has the condition 
that vertex operators are annihilated by $b_0$ (where $b_0$ belongs to $b_{zz}$). 
We have recently shown that the concept of grading is nothing else that 
the ``pure ghost number'' of homological perturbation theory 
\lref\homo{
M.~Henneaux and C.~Teitelboim,
{\it Quantization Of Gauge Systems,}
{\it  Princeton, USA: Univ. Pr. (1992) 520 p}. 
Chap. 8.\semi
J.~M.~Fisch, M.~Henneaux, J.~Stasheff and C.~Teitelboim,
{\it Existence, Uniqueness And Cohomology Of The Classical 
Brst Charge With Ghosts Of Ghosts,}
Commun.\ Math.\ Phys.\  {\bf 120}, 379 (1989).
} \homo. So there is, after all, a deeper geometrical meaning to the ideas we have developed. 


\newsec{Our program applied to Yang-Mills theory}

The program of determining the theory by starting from a suitable set of constraints $d_{z\a}$ 
and a free action (for $x^m, \t^\a$ and $d_\a$) leads to a nilpotent BRST charge and a free action 
in the case of the superstring. Since the ideas are new we would like to see them at work in a simpler 
example. We therefore study in this section whether also for standard pure 
Yang-Mills field theory similar ideas can be implemented and what results they lead to. 

We begin with Yang-Mills fields and write the gauge transformations as 
BRST-like transformations by introducing a ghost field $c^a$ for each 
infinitesimal gauge parameter
\eqn\YMa
{s_0 \, A = \nabla c\,, ~~~~~~~   s_0 \, c =0\,.}
The law $s_0 c =0$ corresponds to $s_0 \l^\a$ and 
$s_0 A = \nabla c$ corresponds to $s_0 \t^\a = i \l^\a$. In string theory we have ``brackets'' 
which are the contraction and propagators of conformal field theory.  
To also introduce brackets for $A$ and $c$, we introduce antifields $A^*$ and 
$c^*$ and define the antibracket
\eqn\YMc{
(X,Y) = {\delta_l X \over \delta z^A}  
{\delta_r Y \over \delta z^*_A} -     
{\delta_l X \over \delta z^*_A}{\delta_r Y \over \delta z^A}
}
for any $X,Y$ in the algebra ${\cal A}$ to construct the rest of the terms 
in $s$. We introduce $A^*, c^*$ the conjugate variable 
to $A,c$ such that $(A^a_\m(x),A^{*}_{b \n} (y) ) =\d^a_b \eta_{\m\n}\d^4(x-y)$ 
and $(c^a(x), c^{*}_b(y)) = \d^a_b \d^4(x-y)$. Notice that though the  
fields $A^*, c^*$ are antifields themselves, in the present section we assign 
antifield number zero to them. In addition, we are not taking into account the Yang-Mills 
equation of motion, but we are only discussing the gauge invariant observables and 
not the observables modulo equations of motion. In the following, we will use 
the antifields as conjugate momenta. The relation between antibracket \YMc~and Poisson 
bracket has been extensively discussed in the literature and we refer to 
\lref\TroostCU{
W.~Troost, P.~van Nieuwenhuizen and A.~Van Proeyen,
{\it Anomalies And The Batalin-Vilkovisky Lagrangian Formalism,}
Nucl.\ Phys.\ B {\bf 333}, 727 (1990).
} \TroostCU~and 
\lref\BarnichMR{
G.~Barnich and M.~Henneaux,
{\it Isomorphisms between the Batalin-Vilkovisky antibracket and the Poisson bracket,}
J.\ Math.\ Phys.\  {\bf 37}, 5273 (1996)
[arXiv:hep-th/9601124].
} \BarnichMR. 

The transformation laws in \YMa~are generated by $S_0 = - \int A^* \nabla c$. 
This corresponds to $Q_0 = \int i \l^\a d_\a$.
The symmetry in \YMa~is not the BRST symmetry because it is not nilpotent $s^2_0 A = - {1\over 2} \nabla [c,c]$ 
where $[\cdot,\cdot]$ is the Lie algebra bracket. However, we can apply again the ideas of 
homological perturbation theory to impose $[c,c] = 0$ as a constraint. This resembles the pure spinor constraint \pure. 
The constraint $[c,c] =0$ is an abelian first 
class constraint and it generates the gauge transformations $\Delta_{\e} c^* = ( {1\over 2}\e \cdot [c,c], c^*) 
= [\e, c]$ where $\e$ is a vector in the adjoint representation and $\cdot$ is the trace operation. 
Finally, the square of the $s_0$ transformations of the fields gives (with $\nabla c = d\, c - [A,c]$)
\eqn\YMaa{
s^2_0 A = \nabla\left( -{1\over 2} [c,c] \right) \,, ~~~~
s^2_0 A^* =\left[ -{1\over 2} [c,c], A^* \right] \,, ~~~~
s^2_0 c =0\,, ~~~~
s^2_0 c^* = \Delta_{\nabla A^*} A^*\,, 
}
which shows that $s_0$ is nilpotent on the 
surface of the constraints modulo gauge transformations.
We introduce a new anticommuting field $\eta^*$ and a differential $\d$ such that 
$\delta$ maps $\eta^*$ into the constraint, and $\delta$ has antifield number 
$\af(\d) =-1$, and $\af(A^*) = \af(c^*) =0$. 
\eqn\YMb
{\d \eta^* = -{1\over 2} [c,c]\,, ~~~~ \d \, A =0\,, ~~~~ \d \, c = 0\,,}
$$
\af(\eta^*) = 1\,, ~~~~ \af(A) = 0\,, ~~~~~ \af(c) =0\,.
$$
We then define the pure ghost number $\pg$ as the sum of the antifield 
number and the ghost number. 
It is easy to check that $\pg(\eta^*) =0$. Applying the theorem of HPT, the two 
operations can be merged in only one nilpotent $s = \d + s_0 + \dots$ since 
the BRST-like transformation $s_0$ is nilpotent modulo $\d$-exact terms. 

A simple exercise shows that 
\eqn\YMd
{ 
\d \,  \eta^* = -{1\over 2} [c,c]\,, ~~~
s_0 \,  \eta^* = - [\eta^*,c]\,, ~~~
s_1 \,  \eta^* = - {1\over 2} [\eta^*,\eta^*]\,,
}
$$
\d \, \eta = 0\,,~~~
s_0 \, \eta = -[\eta,c]\,,~~~
s_1 \, \eta = [\eta,\eta^*] - \nabla A^* \,,~~~
$$
$$
\d \, A^* = 0\,, ~~~
s_0 \, A^* = [c,A^*] \,, ~~~
s_1 \, A^* = [\eta^*,A^*] \,, ~~~
$$
$$
\d\, A = 0\,, ~~~
s_0 \, A = \nabla c\,, ~~~
s_1 \, A = \nabla \eta^* \,, ~~~
$$
$$
\d \, c^* = [c, \eta]\,, ~~~ 
s_0 \, c^* = -\nabla A^* + [\eta^*, \eta] \,, ~~~
s_1 \, c^* = 0 \,, 
$$
$$
\d \, c = 0\,,  ~~~
s_0 \, c = 0\,, ~~~
s_1 \, c = 0\,.
$$
As we already recalled, the construction of the BRST charge, 
which contains both the Koszul-Tate differential $\d$ and the 
BRST-like differential $s_0$, is unique up to a (anti) 
canonical\foot{Anticanonical 
transformations are generated by the antibracket 
$\phi \rightarrow \phi' = \phi +  ({\cal F}, \phi)$ where ${\cal F}$ is a 
fermionic generator.} 
transformation, for example a field redefinition. If 
we shift $\eta^*$ with the ghost field and we 
rename this field $C$ (and in the same way $\eta \equiv C^*$), 
we find out that these transformations can be generated by 
$s X = (S_{\af}, X)$ where $S_{af}$ is 
\eqn\YMe
{
S_{\af} = 
\int d^4x \left( A^* \nabla C + {1\over 2} C^* [C,C] \right)\,.
}
The Lagrangian $S_{af}$ is clearly the usual antifield 
dependent terms of the Yang-Mills Lagrangian. Finally, one can study 
the cohomology of the BRST operator $s$ and one easily finds out 
that the cohomology coincides with the gauge invariant observables of YM theory. 

Notice that by means of the redefinition, we cannot use the 
antifield number to select the resolution of the Koszul-Tate $\d$ any longer. 
Fortunately, in the present case it easy to study the cohomology $H(s)$ directly. 
In addition, the antifield number is protected (it cannot be too negative!!) because 
it is equal to the ghost number.
 
\newsec{Acknowledgments}  

This work was done in part at the Ecole Normale Superieure at Paris whose support we gratefully 
ackwoledge. In addition, we were partly funded by NSF Grants PHY-0098527.


\newsec{Appendix A: Majorana and Weyl spinors in $D=(9,1)$.}

In $D=(9,1)$ dimensions, we 
use ten real $D=(9,1)$ Dirac-matrices $\Gamma^m = \{ I \otimes (i \tau_2), 
\sigma^\m \otimes \tau_1, \chi \otimes \tau_1\}$ 
where $m=0,\dots,9$ and $\m = 1,\dots,8$. The $\sigma^\m$ are eight real symmetric $16 \times 16$ 
off-diagonal Dirac matrices for $D= (8,0)$, while $\chi$ is the real
$16 \times 16$ diagonal chirality matrix in $D=8$\foot{The 8 real $16 \times 16$  matrices
of $D=(8,0)$ can be obtained from a set of 7 pure imaginary $8 \times 8$ 
matrices $\l^i$ for $D=(7,0)$ as follows $\sigma^\m = \{ \l^i \otimes \s_2, I_{8\times8} \otimes \s_1\}$. The 
seven $8 \times 8$ matrices $\l^i$ themselves can be obtained from the 
representation $\g^k = \sigma^k \otimes \tau^2, \g^4 = 1\otimes \tau^1$, and 
$\g^5 = 1\otimes \tau^3$ for $D=(3,1)$ with real symmetric matrices $\gamma^2,\gamma^4,\gamma^5$ and 
imaginary antisymmetric $\gamma^1,\gamma^3$ as follows
$$
\l^i = \{ \g^2 \otimes \s_2, \g^4\otimes \s_2 \otimes \g^5 \otimes \s_2, 
\g^1 \otimes 1, \g^3 \otimes 1, i \, \g^2 \g^4 \g^5 \otimes \s_1, i \g^2 \g^4 \g^5 \otimes \s_3\}.
$$
}. 
So $\chi=\s_1, \dots, \s_8$, $\chi^T = \chi$ and $\chi^2 = 1$. 
The chirality matrix in $D=(9,1)$ is then $I \otimes \tau_3$ and the $D=(9,1)$  
charge conjugation matrix $C$, satisfying  
$C \, \Gamma^m = - \Gamma^{m, T} C$, is numerically equal to $C= \Gamma^0$. If one uses spinors 
$\Psi^T = ( \l_L, \zeta_R)$ with spinor indices $\l_L^\a$ and $\zeta_{R,\dot{\b}}$, 
the index structure of the Dirac matrices, the charge conjugation matrix $C$, 
satisfying $C \G^m = - \G^{m, T} C$ is numerically equal to $\G^0$, and 
the chirality matrix $\G_\# \equiv \G^0 \G^1 \dots \G^9 = I_{16\times 16} \otimes \s_3$ 
is as follows
$$ \Gamma^m = \left( \eqalign{& 0 ~~~~~~~ (\sigma^m)^{\a \dot{\b}} \cr 
 & (\tilde\sigma^m)_{\dot{\b} \g} ~~~ 0 }\right)\,, ~~~
C = \left( \eqalign{& 0 ~~~~~  c_{\a}^{~\dot{\b}} \cr 
 & c^{\dot{\b}}_{~ \g} ~~~ 0 }\right)\,, 
~~~
\G_\# = \left( \eqalign{& I_{16\times16}  ~~~~~  0 \cr 
 & 0 ~~~~~  -I_{16\times16} }\right)\,, 
$$
 where $\sigma^m = \{I, \sigma^\mu, \chi \}$ and $\tilde\sigma^m = \{-I, \sigma^\mu, \chi \}$. 
The 
matrices $c_{\a}^{~\dot{\b}}$ and $ c^{\dot{\b}}_{~ \g} $ are
 numerically equal to $I_{16\times 16}$ and  
 $-I_{16\times 16}$, respectively. 
Thus the $\l^\a$ are chiral and the 
$\zeta_{\dot\b}$ are antichiral. This explains the spinorial index structure of the $\G^m$.
The matrices $\g_m$ satisfy 
$\g^m_{\a\b} \g^{n\,\b\g}+ 
\g^n_{\a\b} \g^{m\,\b\g}=2\eta^{mn}\d_\a^\g$ and 
$\g_{_{m\,(\a\b}} \g^{_{m}}_{_{\g)\d}}=0$. The latter relation makes Fierz rearrangements very easy.

In applications we need the matrices $C\G^m$ (for example in \pure). Direct 
matrix multiplication shows that $C\G^m$ is given by 
\eqn\cacB{
\Gamma^m = \left( \eqalign{& (\tilde\sigma^m)_{\a {\b}} ~~~~ 0\cr 
 &0  ~~~~ - (\sigma^m)^{\dot{\b} \dot\a} }\right) \equiv 
\left( \eqalign{& \g^m_{\a {\b}} ~~~~ 0\cr 
 &0  ~~~~ (\g^m)^{{\b} \a} }\right)\,.}
We only use the real $16\times 16$ symmetric matrices $\g^m_{\a\b} = \tilde\s^m_{\a\b}$ 
and $\g^{\dot\a\dot\b}_m = - \s^{m\dot\a\dot\b}$ in the text, and we omit the dots for 
reasons we now explain. 

The Lorentz generators are given by 
\eqn\cacC{
L^{mn} = {1\over 4} ( \G^m \G^n - \G^n \G^m) = 
\left( \eqalign{& {1\over 2} \s^{m, \a\dot\b} \tilde\s^n_{\dot\b\g} - {m \leftrightarrow n} ~~~~~~~~ 0\cr 
 &0  ~~~~~~~~~~~~~~~~~~~~~~~~~~~~~~~~  - {1\over 2} \s^{m, \a\dot\b} \tilde\s^n_{\dot\b\g} - {m \leftrightarrow n} }\right) 
}
Hence the chiral spinors $\l^\a$ and the antichiral $\zeta_{\dot\b}$ form separate representation 
for $SO(9,1)$. These representations are inequivalent because $\s^m$ and $\tilde\s^m$ are equal 
except for $m=0$ where $\s^0=I$ but $\tilde\s^0 =-I$, and there is no matrix $S$ satisfying 
$S \s^\mu = - \s^\m S$ and $S \chi = - \chi S$. (From $S \s^\mu = - \s^\m S$ it follows 
that $S \chi = + \chi S$). We denote these real inequivalent representation by $\underline{16}$ 
and $\underline{16}'$, respectively. 

In $D=(9,1)$ dimensions one cannot raise or lower spinor indices with the charge 
conjugation matrix, because $C$ is off-diagonal. In $D=(3,1)$, on the other hand, $C$ 
is diagonal and is given by 
$C =\left( \eqalign{& \e_{\a {\b}} ~~~~ 0\cr 
 &0  ~~~~ \e^{\dot{\b} \dot\a} }\right)$, and therefore one can raise and lower the indices 
with the charge conjugation matrices $\e^{\a\b}$ and $\e_{\dot\a\dot\b}$. For that 
reason one has in  $D=(3,1)$ four  kinds of spinors $\l^\a, \l_\a, \chi_{\dot\b}$ and $\chi^{\dot\b}$. 
In $D=(9,1)$ dimension one has only spinors $\l^\a$ and $\chi_{\dot\b}$, and thus one may omit the dots on $\chi_{\dot\b}$ 
without causing confusion. 

We conclude that chiral spinors are given by $\l^\a$, antichiral spinors by $\chi_\a$ and in the text we use 
the twenty real symmetric $16\times 16$ matrices $\g^m_{\a\b}$ and $\g^{m, \a\b}$ (omitting again the 
dots in the latter). The usual Fierz rearrangement for 3 chiral spinors becomes then simply 
the statement that $\g^{_m}_{_{\a\b}} \g_{_{m \g\d}}$ vanishes when totally symmetrized in the indices $\a,\b$ and $\g$.


\lref\Mez{
M. Henneaux and L.~Mezincescu, 
{\it A $\sigma$-model interpretation of the 
Green-Schwarz covariant superstring action}, Phys. Lett. 
{\bf B 152}(1985) 340.
}

\newsec{Appendix B: The WZNW term.}

We follow \Mez. 
The WZNW term ${\cal L}_{WZ}$ is proportional to $\e^{\a\b}$ (with 
$\a,\b =0,1$) hence ${\cal L}_{WZ} d^2x$ can be written as a 2-form
\eqn\WWA{
\omega_2 \equiv {\cal L}_{WZ} d^2x\,.
}
Since $\omega_2$ is susy invariant up to a total derivative,
we have 
\eqn\WWB{
\d_\e \omega_2 = d X\,.
}

Define now a 3-form $\omega_3$ as follows: $\omega_3 = d \omega_2$. Then clearly, 
\eqn\WWC{
\d_\e \omega_3 = 0\,, ~~~~~~~ d \omega_3 = 0\,.
}

From $\d_\e \omega_3=0$ it is natural to try to construct 
$\omega_3$ from the susy-invariant 1-forms $\Pi^m = d x^m - i \sum_j \t^j \g^m d \t_j$ and $d \t^i$. 
Lorentz invariance then yields only one possibility 
\eqn\WWD{
\omega_3 = a_{ij} \Pi^m d \t^i \g^m d \t^j\,.
}
where $a_{ij}$ is a real symmetric $N \times N$ matrix. We diagonalize $a_{ij}$ by a real orthogonal 
transformation (which leaves $\Pi^m$, and thus ${\cal L}_1$, invariant). Then 
$d \omega_3 = -i \left(\sum_i d \t^i \g^m d \t^i \right) \left( \sum_k a_k d \t^k \g^m d \t^k  \right)$.  
In $d \omega_3$ the direct terms cancel due to the 
standard identity $\g^m d\t^1 (d\t^1 \g_m d\t^1) =0$, while the cross-terms cancel only if 
$N=2$ and if the diagonal matrix $a_{ij}$ has entries $(+1,-1)$. Hence
\eqn\WWD{
\omega_3 = -i \Pi^m \left( d \t^1 \g_m d \t^1 - d\t^2 \g^m d\t^2	\right)\,.
}
Using that $\omega_3 = d \omega_2$, 
we find the WZNW term up to an overall constant 
\eqn\WWE{
{\cal L}_{WZ} = - {1\over \pi} \e^{\m\n} 
\left[ i 
\p_\m x^m ( \t^1 \g_m \p_\n \t^1 - \t^2 \g_m \p_\n \t^2 ) + \t^1 \g_m \p_\m \t^1  \t^2 \g^m \p_\n \t^2 
\right]\,.
}
Indeed, 
\eqn\WWF{
d ({\cal L}_{WZ} d^2x) \sim - i d x^m 
\left( d \t^1 \g_m d \t^1 - d\t^2 \g^m d\t^2 \right) + 
\left(\t^1 \g_m d \t^1\, d\t^2 \g_m d\t^2 - d \t^1 \g_m d \t^1\, \t^2 \g^m d\t^2 \right)
}
which is equal to 
\eqn\WWG{
\omega_3 = - i \left(d x^m - \t^1\g_m d \t^1 - \t^2\g^m d\t^2 \right) \left( d\t^1\g_m d\t^1 - d\t^2\g_m d\t^2 \right)\,.
}

Note that the WZNW term is antisymmetric in $\t^1$ and $\t^2$ while ${\cal L}_1$
is symmetric. Only the sum of ${\cal L}_1$ and ${\cal L}_{WZ}$ is $\k$-invariant, up to a total 
derivative. The $\kappa$-transformation rules for $x^m$ read $\d_\k x^m = - \e^i \g^m \d_\k \t^i$ with the opposite 
sign to the susy rule. The expression for $\d_\k \t^\a$ and $\d_\k \sqrt{-h} h^{\m\nu}$ are complicated, involving self-dual 
and antiselfdual anticommuting gauge parameters with 3 indices, but we do not need them. We begin 
with the BRST law $s\, \t^\a = i \l^\a$ where $\l^\a$ is an unconstrained ghost field, but the precise 
classical action to which this corresponds is not know at the present. That does not matter as long as 
we can construct the complete quantum theory, although knowledge of the classical action might clarify the 
results obtained at the quantum level. 

For the open string one has the following boundary conditions at $\s=0,\pi$
\eqn\WWH{
\t^{1 i} = \t^{2 i}\,, ~~~~
\e^{1 i} = \e^{2 i}\,, ~~~~
h^{\s \b} \p_\b x^m =0\,, ~~~~
\k^{1 i}_t = k^{2 i}_t\,. 
}


\newsec{Appendix C: A useful identity for the superstring} 
 
The superstring action is given by 
\eqn\GS{
{\cal L} = - {1\over 2 \pi}  \, \eta_{mn} \, \Pi^m_\mu \Pi^{n\m}  - {\cal L}_{WZ} \,,
} 
$$
{\cal L}_{WZ} = - {1\over \pi} \e^{\m\n} 
\left[i  
\p_\m x^m ( \t^1 \g_m \p_\n \t^1 - \t^2 \g_m \p_\n \t^2 ) + \t^1 \g_m \p_\m \t^1  \t^2 \g^m \p_\n \t^2 
\right]
$$
where $\Pi^m_\mu$ is given in \susyB. 
For definiteness we choose $\e^{01} = 1$ and $\eta^{\m\n}$ as well as 
$\eta^{mn}$ have $\eta^{00} = -1$. This action is real. 

By just writing out all the term, the action can be re-written with chiral 
derivatives 
\eqn\GSA{
- \pi {\cal L} = \eta_{mn} \, \p x^m \bar\p x^n 
- \p x^m \t^1 \g_m \bar\p \t^1 
- \bar\p x^m \t^2 \g_m \p \t^2 
}
$$
+ {1\over 2} (\t^1 \g^m \bar\p \t^1)( \t^1 \g_m \p \t^1 + \t^2 \g_m \p \t^2) 
+ 
{1\over 2} (\t^2 \g^m \p \t^2)( \t^1 \g_m \bar\p \t^1 + \t^2 \g_m \bar\p \t^2) 
$$
with $\p = \p_\s - \p_t$ and $\bar\p = \p_\s + \p_t$. 

Except for the purely bosonic terms, all terms involve either $\p\t$ or $\bar\p\t$. Hence 
we can write the action as 
\eqn\GSB{
- \pi {\cal L} = \eta_{mn} \, \p x^m \bar\p x^n + (p_{1 \a})_{\rm Sol}  \bar \p \,\t^{1 \a} + 
(p_{2 \a})_{\rm Sol} \p \, \t^{2\a}  
}
where $(p_{i \a})_{\rm Sol}$ are complicated composite expressions. We restrict 
ourselves to the left-moving sector, setting $\t^2 = p_2 =0$. 

We can then 
also write the action with independent $p_{i \a}$ if we impose the constraint that 
$d_{i \a}  \equiv p_{i \a} - (p_{i \a})_{\rm Sol}$ vanishes. Finally, 
the complete expressions are given by
\eqn\GSC{
d_{1 \a} = p_{1 \a} +  \p x^m \t^1 \g_m \bar\p \t^1 - 
{1\over 2} (\t^1 \g^m \bar\p \t^1)( \t^1 \g_m \p \t^1 + \t^2 \g_m \p \t^2) \,, 
}
$$
d_{2 \a} = p_{2 \a} +  \p x^m \t^2 \g_m \bar\p \t^2 - 
{1\over 2} (\t^2 \g^m \bar\p \t^2)( \t^1 \g_m \p \t^1 + \t^2 \g_m \p \t^2) \,. 
$$
In the text we work with the free action with independent fields $p_{i \a}$. The 
$d_{i \a}$ are transferred to the BRST charge where they are multiplied by the independent 
unconstrained real chiral commuting spinors $\l^\a$.  To make use of the calculation technique 
of conformal field theory, we made a Wick rotation $t \rightarrow - i \tau$, $\p_t \rightarrow 
+ i \p_\tau$ and $\p = \p_\s - \p_\tau \rightarrow \p = \p_\s - i \p_\tau$ and analogously for $\bar\p$. 
We also restrict ourselves to only one sector with $\t = \t^1$ and $d_{\a} = d_{1\a}$, by setting $\t^2 = 0$. 
For a treatment which describes both sectors, we refer to \GrassiXF. 


\newsec{Appendix D: Solution of the pure spinor constraints.}

In this appendix we discuss a solution of the constraint that the 
chiral spinors $\l$ are pure spinors. The equation to be solved 
reads 
\eqn\puu{
\l^\a \g^m_{\a\b} \l^\b =0\,,
} 
where $\l^\a$ are complex chiral (16-component) spinors. We shall decompose $\l$ w.r.t. a non-compact 
version of the $SU(5)$ subgroup of $SO(9,1)$ as 
$|\l\rangle = \l_+ |0\rangle + {1\over 2!} \l_{ij} a^i a^j |0\rangle +  {1\over 4!} \l_{jklm} a^j a^k  a^l a^m |0\rangle$. 
This decomposition corresponds to $\underline{16} = \underline{1} + \underline{10} + \underline{5}^*$. Then 
we shall show that the constraints express the $\underline{5}^*$ in terms of the $\underline{1}$ and $\underline{10}$. 
Hence there are 11 independent complex components in $\l$. We shall prove that $\l$ is complex and not a Majorana 
spinor, so $\bar\l_D \equiv \l^\dagger i \g^0$ differs from $\bar\l_M = \l^T C$. 
(Recall that a Majorana spinor is defined by the 
condition $\bar\l_D = \bar\l_M$). 

The Dirac matrices in $D=(9,1)$ dimensions satisfy $\{\G^m, \G^n\} = 2 \eta^{mn}$, where $\eta^{mn}$ is 
diagonal with entries $(-1,+1, \dots, 1)$ for $m,n=0,\dots,9$. We combine them into 5 annihilation operators $a_j$
and 5 creation operators $a^j = a^\dagger_j$ as follows 
\eqn\puuA{
a_1 = \half(\G^1 + i \G^2)\,, ~~~~~
a_2 = \half(\G^3 + i \G^4)\,, ~~~\dots~~~~
a_5 = \half(\G^9 -  \G^0)\,.
}
$$
a^1 = \half(\G^1 - i \G^2)\,, ~~~~~
a^2 = \half(\G^3 - i \G^4)\,, ~~~\dots~~~~
a^5 = \half(\G^9  +  \G^0)\,.
$$
Clearly $\{ a_i, a^j \} = \d_i^j$ for $i,j= 1,\dots,5$. We introduce a vacuum $|0 \rangle$ 
with $a_i |0 \rangle = 0$. By acting with one or more $a^j$ on  $|0 \rangle$, we obtain 32 states $| A \rangle$
with $A =1, \dots, 32$. . 
Similarly, we introduce a state $\langle 0|$ which satisfies  $\langle 0|\a^j = 0 $ and we create 32 states $\langle B|$ by acting 
with one or more $a_i$ on $\langle 0|$. We choose the states $\langle B|$ as $| A \rangle^\dagger$. For example, 
if $| A \rangle = a^{i_1} \dots a^{i_k} |0 \rangle $ then $\langle A| = \langle 0| a_{i_k} \dots a_{i_1}$. Then 
$\langle A | B \rangle = \delta^A_{~~B}$.

{\it Lemma 1:} The matrix elements $\langle B| a^j |A\rangle \equiv (\G^j)^B_{~~A}$ and 
$\langle B| a_j |C\rangle \equiv (\G_j)^B_{~~C}$ form a representation 
of the Clifford algebra. 

{\it Proof:} This follows from $\sum |C\rangle  \langle C| = I$. Namely, 
$\sum |C\rangle  \langle C|  =  |0 \rangle  \langle 0| + \sum_i a^i  |0 \rangle  \langle 0| a_i + \dots + 
 a^1\dots a^5  |0 \rangle  \langle 0| a_5 \dots a_1$, where the sum over $C$ runs over the 32 states shown. 
For any state $| A \rangle$ one has $| A \rangle = \sum |C\rangle  \langle C| A \rangle$, because 
$  \langle C| A \rangle = \d^C_{~~A}$ by construction. 

{\it Lemma 2:} The chirality matrix $\G_\# = \G^1\G^2\dots\G^9\G^0$ satisfies $\G_\#^2 = 1$, and  
$\G_\#^\dagger = \G_\#$. It is given by 
\eqn\puuB{
\G_\# = (2 a_1 a^1 -1) \dots ( 2 a_5 a^5 -1) \,.
}

{\it Proof:} $\G_\# = (a^1 + a_1)(a^1 - a_1) \dots (a^5 + a_5) (a^5 - a_5)$ and 
$ (a^1 + a_1)(a^1 - a_1) = (2 a_1 a^1 -1)$. As a check note that $(2 a_1 a^1 -1)^2=1$, and that 
$\{\G_\#, a^1 \} =0$ because $\{  (2 a_1 a^1 -1) , a^1 \} =0$. Similarly $\{\G_\#, a_1\} =0$. Further, 
$\G_\# |0 \rangle =  |0 \rangle$. 

{\it Lemma 3:} $\langle B| a^j |C\rangle = \langle C| a_j |B\rangle =$ real. 

{\it Proof:} This follows from the fact that one obtains the second matrix element 
from the first by left-right reflection, and from the fact that the anticommutation 
relations have the same symmetry and are real: $\{a^k, a_l\} =\{a_l, a^k\} = \d^k_{~l}$. 

{\it Lemma 4:} The matrix representation of $\G^1,\G^3,\G^5,\G^7, \G^9$ is real 
and symmetric while that of $\G^2,\G^4,\G^6, \G^8$ and $\G^0$ is purely imaginary and antisymmetric. 

{\it Proof:} $\langle A| a^j \pm a_j |B \rangle = \langle B| \pm a^j + a_j |A\rangle$. 

{\it Lemma 5:} The charge conjugation matrix $C$, defined by $C \G^m = - \G^{m,T} C$ is given by 
$C = - \G^2 \G^4 \G^6 \G^8 \G^0 = (a_1 - a^1) (a_2 -a^2) \dots (a_5 -a^5)$. The minus sign is added for later 
convenience. 

{\it Proof:} $\G^1,\G^3,\G^5,\G^7, \G^9$ anticommute with $C$, while 
$\G^2,\G^4,\G^6, \G^8,\G^0$ commute with $C$, the former are symmetric while the latter are antisymmetric. 

{\it Theorem I:} A chiral spinor $\l$ can be expanded as follows 
\eqn\puuD{
|\l \rangle = \l_+  |0\rangle +  
{1\over 2!} \l_{ij} a^j a^i |0\rangle +  {1\over 4!} \l^i \e_{ijklm} a^j a^k  a^l a^m |0\rangle\,. 
}

{\it Proof:} $\G_\# |0\rangle = |0\rangle$; hence $\G_\# |\l \rangle = |\l \rangle$. The 16 non-vanishing 
components of $|\l \rangle$ are the projections of the ket $|\l \rangle$ onto the corresponding 16 bras: 
in particular
\eqn\puuE{
\l_+ = \langle 0| \l \rangle = \langle \l|0 \rangle \,, 
~~~~~
\l_{ij} = {1\over 2!} \langle 0|a_i a_j| \l \rangle = {1\over 2!} \langle \l| a^j a^i|0 \rangle \,, 
}
$$
 \l^i =  {1\over 4!} \e^{ijklm} \langle 0| a_j a_k  a_l a_m | \l \rangle = 
{1\over 4!} \e^{ijklm} \langle \l| a^j a^k  a^l a^m |  0 \rangle\,.
$$

We are now ready to solve the ten constraints $\bar\l^\a \G^m_{\a\b} \l^b =0$. These 
relations are equivalent to the five constraints $\l^T C a^j \l = 0$ and the five other 
constraints $\l^T C a_j \l = 0$. They can be rewritten as follows 
\eqn\puuF{
\langle \l| C a^j | \l \rangle = 0\,, ~~~~~
\langle \l| C a_j | \l \rangle \,. 
}

{\it Theorem II:} $\langle A| C |B\rangle \neq 0$ iff $A^\dagger B$ is 
proportional to precisely $a^1a^2 a^3 a^4 a^5 |0\rangle$. 

{\it Proof:} $a_j C = - C a^j$ and $a^j C = - C a_j $, Further $C |0\rangle = - a^1a^2 a^3 a^4 a^5 |0\rangle$ 
and $\langle 0| C =  \langle 0|  a^5a^4 a^3 a^2 a^1$. Pulling all $a_j$ in $\langle A|$ to the right of 
$C$, we obtain, up to an overall sign, $\langle 0| C A^\dagger B |0 \rangle$ and this is only non-vanishing if all $a^k$ in 
$A^\dagger B$ match the $a_k$ in $\langle 0|C$. It follows that $ \langle 0| C a^1a^2 a^3 a^4 a^5 |0\rangle = 1$. 

{\it First set of constraints}
\eqn\puuG{
\langle \l| C a^{i_0} | \l \rangle = \langle 0|C \left( \l_+ + {1\over 2} \l_{ij} a^i a^j + 
{1\over 4!} \l^i a^j a^k a^l a^m \e_{ijklm} \right) a^{i_0} \l \rangle 
 }
$$
=2 \left( \l_+ \l^{i_0} + {1\over 4!}  \e^{i_0 jklm} \l_{jk} \l_{lm} \right)\,.
$$

{\it Second set of constraints}
\eqn\puuH{
\langle \l| C a_{i_0} | \l \rangle = \langle 0|C \left( \l_+ + {1\over 2} \l_{ij} a^i a^j + 
{1\over 4!} \l^i a^j a^k a^l a^m \e_{ijklm} \right) a_{i_0} \l \rangle 
 }
$$
=- 2 \l_{i_0 j} \l^j \,.
$$

{\it Main Result:} The solution of the first set of constraints 
$\l_+ \l^{i} + {1\over 4!}  \e^{i jklm} \l_{jk} \l_{lm} = 0$ is given by 
\eqn\puuI{
\l^i = - {1\over 4! \l_+}  \e^{i jklm} \l_{jk} \l_{lm}\,. 
}
The solution automatically satisfies the second set of constraints because 
\eqn\puuK{
\l^i \l_{in} = \e^{ijklm} \l_{jk} \l_{lm} \l_{in} = 0\,.
}

{\it Proof:} A totally antisymmetric tensor with $6$ indices in 5 dimensions vanishes. Hence 
$\l^i \l_{in}$ is equal to a sum of 5 terms, due to exchange $n$ with $j,k,l,m$ and $i$, respectively. Interchanging 
$n$ with $i$ yields minus the original tensor, but also interchanging $n$ with $j,k,l$ and $m$ yields each time minus the original 
expression. Hence the expression vanishes. 

{\it Comment 1:} The fact that a pure chiral spinor contains 
11 independent complex components leads to a vanishing 
central charge in Berkovits' approach with variables $x^m, \t^\a$ and the 
conjugate momentum $p_\a$, and $\l^\a$ with conjugate momentum $p_{(\l)\a}$: 
$c = + 10_{x} - 2 \times 16_{\t p} + 2 \times 11_{\l, p_{\l}} = 0$. 
In our approach we have $16$ independent real component in $\l^\a$ and 16 conjugate 
momenta $p_{(\l)\a}$ with $\a =1,\dots,16$. Also in our case $c=0$, but there are more ghosts, and 
there is nowhere a decomposition w.r.t. a subgroup of $SO(9,1)$. 

{\it Comment 2:}  
In the decomposition in Theorem I, one can choose all $\l$'s to be real, and $\l^i$ to be expressed in terms 
of $\l_+$ and $\l_{ij}$ as in \puuI. Then $\l$ is a real chiral spinor. However, the Dirac matrices are complex, 
so under a Lorentz transformation $\l$ becomes complex in a general Lorentz frame. 


\footatend\vfill\supereject\immediate\closeout\rfile\writestoppt
\baselineskip=14pt\centerline{{\bf References}}\bigskip{\frenchspacing%
\parindent=20pt\escapechar=` \input refs.tmp\vfill\eject}\nonfrenchspacing  
  
\bye